# Large and Robust Charge-to-Spin Conversion in Sputtered Conductive WTe$_x$ with Disorder

Xiang Li[1,2*i], Peng Li[3], Vincent D.-H. Hou[4], Mahendra DC[1], Chih-Hung Nien[4], Fen Xue[1], Di Yi[3], Chong Bi[2], Chien-Min Lee[4], Shy-Jay Lin[4], Wilman Tsai[4], Yuri Suzuki[3], and Shan X. Wang[1,2**]

[1]Department of Materials Science and Engineering, Stanford University, Stanford, California 94305, USA

[2]Department of Electrical Engineering, Stanford University, Stanford, California 94305, USA

[3]Department of Applied Physics, Stanford University, Stanford, California 94305, USA

[4]Corporate Research, Taiwan Semiconductor Manufacturing Company, Hsinchu, Taiwan

Corresponding Authors: *xiangshaunli@gmail.com, **sxwang@stanford.edu

[i] Lead Contact: Xiang Li, xiangshaunli@gmail.com

## Summary

Topological materials with large spin-orbit coupling and immunity to disorder-induced symmetry breaking show great promise for efficiently converting charge to spin. Here, we report that long-range disordered sputtered $WTe_x$ thin films exhibit local chemical and structural order as those of Weyl semimetal $WTe_2$ and conduction behavior that is consistent with semi-metallic Weyl fermion. We find large charge-to-spin conversion properties and electrical conductivity in thermally annealed sputtered $WTe_x$ films that are comparable with those in crystalline $WTe_2$ flakes. Besides, the strength of unidirectional spin Hall magnetoresistance in annealed $WTe_x$/Mo/CoFeB heterostructure is 5 to 20 times larger than typical SOT layer/ferromagnet heterostructures reported at room temperature. We further demonstrate room temperature damping-like SOT-driven magnetization switching of in-plane magnetized CoFeB. These large charge-to-spin conversion properties that are robust in the presence of long-range disorder and thermal annealing pave the way for industrial application of a new class of sputtered semimetals.

**Introduction**

Recently, topological insulators (TIs) [1,2] emerge as a promising class of materials for converting charge current to spin-polarized current. Owing to their strong spin-orbit coupling and spin-momentum locked surface states [3,4], TIs promise large charge-to-spin conversion efficiency ($\xi_{ST}$) and energy-efficient switching of an adjacent ferromagnetic metal (FM) via spin-orbit torque (SOT). However, bulk states that dominate electron conduction in some TIs such as $Bi_2Se_3$ might not convert charge to spin as efficiently as surface states.[5-7] Also, when interfacing TIs with FM in a magnetic tunnel junction (MTJ) configuration, magnetic impurities diffused out of the FM layer may drive the TI into a trivial state,[8] making the TI/FM interface vulnerable to annealing or thermal effects. Lastly, for practical applications such as SOT-Magnetic Random Access Memory (MRAM), as the tunneling magnetoresistance of an MTJ is rather limited to several hundred percent, an insulating SOT write line would significantly degrade the MTJ readout signal, while also increase the parasitic resistance in the write path thereby decreasing the write current a transistor can provide.[9]

In contrast, Weyl semimetals (WSMs) such as type-II WSM $WTe_2$, shows great prospect to overcome the above challenges. First, both Fermi arcs surface states that connect the Weyl points, and the bulk electron and hole pockets are spin-polarized in $WTe_2$. [10] Meanwhile, the Weyl points and spin-momentum-locked surface states persist under broken time reversal symmetry, thus magnetic impurities doping.[11] In addition, compared with the surface states in TIs and type-I WSM with a point-like Fermi surface, a type-II Weyl point appears at the contact of bulk electron and hole pockets.[12] Hence, there is semi-metallic transport in type-II WSM through the bulk electron and hole pockets. This semi-metallic transport leads to their higher conductivity ($\sigma_{xx}$) compared with TIs, [13] therefore smaller parasitic resistance in the MTJ read and write paths, and more efficient use of the current and power provided by the driving transistor.

In the past few years, large $\xi_{ST}$ has been demonstrated in TI/FM heterostructures at room temperature [5,6]. Lately, research also shows a large in-plane damping-like $\xi_{ST}$ in $WTe_2$ [14,15] and low-temperature enhancement of field-like torque when current flows along its *b* axis [16], as well as an out-of-plane damping-like $\xi_{ST}$ when current flows along its *a* axis.[17] However, most of the spintronics studies using topological materials employ crystalline TIs grown by molecular beam epitaxy (MBE) [5, 6, 18] and exfoliated WSM flakes from bulk crystals [14, 16, 17], which are not suitable for large-scale industrial manufacturing. Therefore, it is highly desirable to use industrial deposition techniques, such as sputtering to grow these novel topological materials.

Recently, sputtered $Bi_xSe_{(1-x)}$ films with nanoscale crystallites have shown high $\xi_{ST}$ up to 18 [19], though their high resistivity up to 13000 $\mu\Omega\ cm$ severely hinders the use of sputtered TIs in MRAM applications. Hence, there is a strong need for sputtered conductive topological materials with large $\xi_{ST}$ and $\sigma_s^{eff}$ values for integration into the metallic MTJ system. Meanwhile, polycrystalline $Bi_2Se_3$ prepared by pulsed laser deposition have exhibited topological surface states [20]. While even in amorphous $Bi_2Se_3$ prepared by thermal evaporation, researchers have observed spin-polarized surface states using angle-resolved photoemission spectroscopy [21]. Though amorphous materials do not host Bloch states due to a lack of long-range order, band structures in topologically disordered amorphous semiconductors can be calculated in theory based on the local atomic environment.[22, 23] In fact, the electronic properties of solids are dominated by short-range order while the long-range order is responsible only for the finer band structures.[22] Recently, several theoretical works have shown that electrons hopping among randomly distributed atoms still present topologically protected edge states.[24, 25] Moreover, a recent theoretical work shows a topological metal phase in 3D amorphous metals exhibiting surface states.[26] Nevertheless, no work has experimentally investigated the effect of disorder in topological semimetals on their electronic and spintronic properties.

Here, we report large and robust charge-to-spin conversion properties in sputtered disordered $WTe_x$. We first confirm our sputtered $WTe_x$ materials show similar local chemical and structural order as crystalline $WTe_2$ at room temperature. Through low-temperature magnetoresistance and temperature-dependent resistance measurements, we find that $WTe_x$ films exhibit behaviors that are consistent with Weyl fermions conduction. An increase of long-range disorder with $WTe_x$ thickness drives a crossover from weak anti-localization to weak localization, a decrease in bulk semi-metallic conduction, and an increase in bulk variable range hopping transport. Using spin-torque ferromagnetic resonance (ST-FMR), we find increasing $\xi_{ST}$ and decreasing effective spin Hall conductivity ($\sigma_s^{eff}$) with $WTe_x$ thickness, suggesting a bulk spin Hall effect origin of the SOT and the material transitioning into dirty metal regime. We find in thermally annealed $WTe_x/Mo/CoFeB$ heterostructure that sputtered $WTe_x$ films show $\sigma_s^{eff}$ of

$0.98 \times 10^5 (\hbar/2e)\ \Omega^{-1} m^{-1}$, $\xi_{ST}$ of 0.4, and resistivity of 435 $\mu\Omega\ cm$ at room temperature, which are comparable with those reported in crystalline WTe$_2$ flakes. We also find 5-20 times larger unidirectional spin Hall magnetoresistance (USMR) signal in the annealed heterostructure than existing FM/heavy metal (HM) or FM/TI bilayers. We further demonstrate room-temperature damping-like SOT-driven magnetization switching of in-plane magnetized CoFeB using a simple planar USMR device geometry. Together these results indicate that sputtered WTe$_x$ films are a promising candidate for spintronics applications based on metallic MTJs as well as planar devices based on SOT layer/FM interfaces.

## Results

### Material Properties of WTe$_x$ Thin Films

We first confirm the sputtered WTe$_x$ films exhibit local chemical and structural order as crystalline WTe$_2$ via Raman spectroscopy and atomic-scale structural and chemical characterizations. Raman spectroscopy is a widely used technique to identify materials based on their unique phonon vibrational mode fingerprints determined by crystal structure and chemical bonding. As shown in Figure 1**a**, we observe Raman modes $B_1^{10}, A_2^4, A_1^9, A_1^6, A_1^5, A_1^2$ in uncapped single layer WTe$_x$ samples with different thicknesses ($t_{WTe_x}$). As $t_{WTe_x}$ increases, $A_1^5$ and $A_1^2$ Raman modes show smaller widths, while the $A_1^2$ peak frequency decreases (Figure S1), which are consistent with the reports on bulk and flaked WTe$_2$ films. [27] Meanwhile for the 5 nm thick sample, only $A_1^5, A_1^2$ modes are visible, which is similar as reported in monolayer exfoliated WTe$_2$ flakes. [27] We next employ X-ray Photoelectron Spectroscopy (XPS) to confirm that the chemical bonding state of Te atoms in our WTe$_x$ material is the same as that in WTe$_2$. As shown in Figure 1**b**, the binding energy of Te 3d$_{5/2}$ electron is consistent with WTe$_2$ state reported in exfoliated flakes (572.7 eV), compared with metallic Te (573.0 eV) and TeO$_x$ (576.3 eV). [28] Further, as shown in Figure S2**b-c**, a high-resolution scanning transmission electron dark-field microscopy (HR-STEM) image reveals a small crystalline cluster embedded in amorphous WTe$_x$ films.

Then we study the evolution of long-range disorder with the WTe$_x$ films thickness. In Figure 1**a**, two groups of Raman modes co-exist in WTe$_x$ films, namely a first group of

$B_1^{10}, A_1^9, A_1^6, A_1^5, A_1^2$ modes and a second group of $B_1^{10}, A_2^4, A_1^5, A_1^2$ modes, which appear at different locations on the same 58 nm thick WTe$_x$ sample (more data in Figure S1). This first group appears when the laser is aligned along the $b$ axis of bulk WTe$_2$ crystal, and the second group along the $c$ axis.[27] The data suggest that as $t_{WTe_x}$ increases, the WTe$_x$ films segregate into small crystalline WTe$_2$ clusters with their $b$ or $c$ axis aligned normal. Further, as the STEM images Figure 1**c-d** show, the 5 nm thick WTe$_x$ film exhibits a mixture of sub-nm features of bright and dark contrast, while in the 58 nm sample, ~10-nm features of segregated dark clusters surrounded by bright regions emerge. Using energy-dispersive X-ray spectroscopy (EDS) shown in Figure S3, we find that the bright regions are W-rich, while the dark regions are Te-rich. Thus, we conclude that as $t_{WTe_x}$ increases, long-range segregation disorder intensifies.

**Electronic Properties of WTe$_x$ Thin Films**

To study the electronic properties of the WTe$_x$ thin films with different thickness thus long-range disorder, we measure magnetoresistance (MR) and temperature-dependent resistance. We find the increase of disorder and electron-phonon interaction drives a crossover from weak anti-localization (WAL) in the 5 nm thick WTe$_x$ to weak localization (WL) in the 41 nm thick WTe$_x$. The WAL (WL) effect describes the destructive (constructive) quantum interference of electron waves going around a self-intersecting path in opposite directions in a disordered electronic system with spin-orbit coupling (SOC).[29] As shown in Figure 2**a**, the 5 nm thick WTe$_x$ film shows a positive MR with a cusp-shaped perpendicular magnetic field ($B_z$) dependence from 4 K to 50 K. In contrast, we find a negative MR in 41 nm thick WTe$_x$ devices from 4 K to 20 K. More analysis shows that the WAL behavior is consistent with thin WTe$_2$ flakes with strong disorder but differs from WTe$_2$ crystals or thick flakes (See Supplementary Note S4).

We find that quantum transport in both thin and thick WTe$_x$ samples from 4 K to 50 K can be well described by Weyl fermion, which characterizes the linear band dispersion near a Weyl point. This is possible because the local chemical and structural order in sputtered WTe$_x$ materials discussed in Figure 1 still preserves the essential band structure like that in crystalline WTe$_2$.[22] As shown in Figure 2**c**, we first fit the 5 nm thick WTe$_x$ MR curves using the simplified Hikami-Larkin-Nagaoka (HLN) equation, which describes the WAL correction to conductivity $\Delta\sigma_{xx} = \sigma_{xx}(B_z) - \sigma_{xx}(0)$ under $B_z$ in a quasi-2D system with strong SOC [29]: $\Delta\sigma_{xx} \cong$

$-\frac{\alpha_{HLN} e^2}{2\pi h}\left[\psi\left(\frac{1}{2}+\frac{B_\phi}{B_z}\right)-\ln\left(\frac{B_\phi}{B_z}\right)\right]$, where $\psi$ is the digamma function, $B_\phi$ is the phase coherence characteristic field, and $\alpha_{HLN}$ is 0.5 for WAL and -1 for WL. The linear dependence of $B_\phi$ on temperature in Figure 2**d** indicates that electron-electron interactions cause the dephasing of electron waves.[30] Similar $\alpha_{HLN}$ decrease as temperature increases as well as $\alpha_{HLN}$ values between 0 and 0.5 have been observed in Dirac semimetal $Cd_3As_2$ films.[31, 32] We further confirm the 2D nature of the 5 nm thick $WTe_x$ film in Supplementary Note S5. Moreover, as shown in Figure 2**b**, the change of sheet conductance $\Delta\sigma_{xx} = \sigma_{xx}(T) - \sigma_{xx}(4K)$ as a function of temperature can be fitted using an equation describing 2D massless Dirac fermions [33], i.e., $\Delta\sigma_{xx} = (c_{ee} - c_{qi})\ln(T)$, where $c_{ee}, c_{qi}$ describe the contributions from interplay of electron-electron interaction and disorder scattering, and quantum interference (WAL in this case), respectively. The positive fitted $c_{ee} - c_{qi}$ value of $1.83\pm0.02$ also suggests a large role of disorder in our films. Note that as in-plane inversion symmetry is broken in the small $WTe_2$ crystallites as shown in Figure S2**b** as well as among randomly ordered $WTe_2$ crystallites in the 5 nm thick amorphous $WTe_x$, a Dirac node separates into two Weyl nodes.[12] Hence, we find the transport properties of the 5 nm thick $WTe_x$ are consistent with quasi-2D massless Weyl fermion with electron-electron interaction.

For the 41 nm thick sample, we fit the change in sheet conductance $\Delta\sigma_{xx}$ as a function of $B_z$ using an equation describing WL and WAL in 3D Weyl semimetal as proposed in [34]: $\Delta\sigma_{xx} = \frac{C_1^{qi} B_z^2 \sqrt{B_z}}{B_\phi^2 + B_z^2} + \frac{C_2^{qi} B_\phi^2 B_z^2}{B_\phi^2 + B_z^2}$, where fitting parameters $C_1^{qi}, C_2^{qi}$ are positive for WL and negative for WAL. As shown in Figure 2**d**, the cubic $B_\phi$ dependence on temperature indicates electron-phonon interaction is the dominating dephasing process, [30] while the non-zero $B_\phi$ value at 0 K suggests surface scattering and impurity also play a role.[31] We also verify the validity of using 3D equation in Supplementary Note S5. Next, we fit the temperature dependence of $\Delta\sigma_{xx}$ using an equation describing 3D massless Weyl fermions, $\Delta\sigma_{xx} = c_{ee}T^{0.5} - c_{qi}T^{p/2}$.[33] The good fit using $p = 3$ further suggests that the 41 nm thick $WTe_x$ transport can be described by 3D Weyl fermion with electron-phonon interaction. The fitted values of $c_{ee} = 2.16\pm0.02$ and $c_{qi} = (1.07\pm0.05)\times 10^{-3}$ indicate a strong interplay between electron-electron interaction and disorder scattering. From theory, this large $c_{ee}$ also leads to a crossover from WAL to WL in WSM.[33] Therefore, the increasing long-range disorder as $t_{WTe_x}$ increases may lead to the dominating electron-phonon interaction and drives the crossover from WAL to WL.

Next, we analyze the temperature dependent resistivity $\rho$ of WTe$_x$ with different thicknesses extracted from as-deposited WTe$_x$/CoFeB bilayers (Supplementary Note S6). First the resistivity of WTe$_x$ at room temperature increases from 160 to 1500 $\mu\Omega\,cm$ as $t_{WTe_x}$ increases from 5 to 58 nm, and all thicknesses show an insulating behavior with a sharp upturn at low temperatures, as shown in Figure 3**a**. This is in sharp contrast to crystalline WTe$_2$ where resistivity increases at small thicknesses due to stronger interface scattering and decreases at low temperatures.[14, 16, 35] Given the enhanced long-range disorder as $t_{WTe_x}$ increases, we postulate the WTe$_x$ material enters dirty metal regime with variable range hopping (VRH) transport behavior [36]. This postulation is corroborated as dirty metal is defined when $\rho$ increases to 100 – 1000 $\mu\Omega\,cm$ [37, 38]. However, a VRH model cannot fit the whole temperature range data (Supplementary Note S6). Considering that the Weyl fermions-like transport behaviors discussed in Figure 2**b** indicates semi-metallic transport, we thus adopt a two-channel model used widely in TIs studies to fit the whole range data as seen in Figure 3**a**. [39, 40] This model consists of one conduction channel of insulating VRH behavior and the other conduction channel of semi-metallic behavior that is independent of temperature [41]: $\frac{1}{\rho} = \frac{1}{\rho_{metal}} + \left[\rho_{VRH} e^{\left(\frac{T_0}{T}\right)^{0.25}}\right]^{-1}$ where $\rho_{metal}$ is the semi-metallic channel resistivity of WTe$_x$, $\rho_{VRH}$ is the VRH channel resistivity of WTe$_x$, and $T_0$ is the characteristic Mott temperature. In Supplementary Note S6, we further confirm the validity of this model. This insulating behavior is also consistent with Figure 2**b** where the large $c_{ee}$ suggests the significant role of disorder. We then calculate the contribution of the semi-metallic channel $G_{semi-metallic\,channel}^{WTe_x}$ to the sum of both semi-metallic and VRH channels $G_{both\,channels}^{WTe_x}$ in Figure 3**b**. The decreasing semi-metallic conduction across the whole temperature range as $t_{WTe_x}$ increases can be attributed the increasing long-range disorder that destructs bulk semi-metallic states into localized hopping states.

It is worth noting that diffusive electron transport behaviors such as WAL and WL are different from Mott-type VRH electron transport. Nevertheless, an intermediate regime that has characteristics of both types of transport has been observed in ultrathin TI films with sheet resistivity ($R_s$) of 10 - 100 $k\Omega/sq$.[42] This experiment also shows suppressed $\alpha_{HLN}$ of 0.01 - 0.1 in this intermediate regime, which agrees with the $\alpha_{HLN}$ of 0.05 - 0.09 in our 5 nm WTe$_x$ sample with $R_s$ of 13 - 37 $k\Omega/sq$. Research has shown WAL and VRH transport behaviors in degraded crystalline WTe$_2$ flakes after exposure to air, consisting of 10 x 10 WTe$_2$ atoms crystalline clusters embedded in an amorphous matrix, [43] which is similar to that in Figure S2**b**. In addition, the room temperature resistivity of the 5 nm WTe$_x$, i.e. 159 $\mu\Omega\ cm$, is smaller than previous reported 460-1400 $\mu\Omega\ cm$ for exfoliated WTe$_2$ flakes from 5L to 17L.[44] Considering that the semi-metallic channel consists of over 90% of conduction at room temperature as shown in Figure 3**b**, we speculate there are additional conduction channels arising in these amorphous WTe$_x$ films compared with crystalline counterparts.

**Charge-to-Spin Conversion in As Deposited WTe$_x$/CoFeB Bilayers**

We then measure large charge-to-spin-conversion efficiency ($\xi_{ST}$) up to 0.5 in the as-deposited WTe$_x$/CoFeB bilayers with varying $t_{WTe_x}$ mentioned above using the spin-torque ferromagnetic resonance (ST-FMR) technique. As illustrated in Figure S7**c**, by flowing GHz RF current through the WTe$_x$/CoFeB bilayers, the oscillating CoFeB anisotropic magnetoresistance driven by damping-like and Oersted-field torques ($\tau_{DL}, \tau_{Oe}$) generates a DC voltage when mixed with the RF current. Based on established methodology (Supplementary Note S7)[45], we fit the mixing voltage $V_{mix}$ using symmetric and asymmetric Lorentzian shapes which originate from $\tau_{DL}$ and $\tau_{Oe}$ respectively as in Figure 4**a**. The extracted $\xi_{ST}$ values for different WTe$_x$ thickness samples are shown in Figure 4**b**, where each $\xi_{ST}$ value is averaged across 6 and 7 GHz measurements. The increasing $\xi_{ST}$ from 0.15 to 0.5 as a function of $t_{WTe_x}$ is comparable with that from the exfoliated WTe$_2$ flake.[14] We also find the effective spin Hall conductivity $\sigma_s^{eff} = (\hbar/2e)\xi_{ST}/\rho_{xx}$ decreases with $t_{WTe_x}$. As shown in Figure 4**b**, this is driven by the much faster increase in $\rho_{xx}$ than the increase in $\xi_{ST}$ as a function of $t_{WTe_x}$.

We next discuss possible causes of this reduction in $\sigma_s^{eff}$ and origins of the SOT in sputtered WTe$_x$. According to spin Hall effect (SHE) theory, [37] $\sigma_s^{eff}$ decreases rapidly when $\sigma_{xx}$ decreases to $10^5 - 10^6\ \Omega^{-1}m^{-1}$, thereby exiting the intrinsic metal regime and entering the dirty metal regime. This is also what happens in our films as shown in Figure 4**c**. We thus further confirm the postulation that our films enter the dirty metal regime as discussed in Figure 3, where the decrease in semi-metallic transport with increasing $t_{WTe_x}$ is also consistent with the decrease of $\sigma_s^{eff}$ with $t_{WTe_x}$. *Ab initio* theory [46] has calculated spin Hall conductivity $\sigma_s$ up to $0.5 \times 10^5 (\hbar/2e)\ \Omega^{-1} m^{-1}$ in WTe$_2$ due to intrinsic spin Hall effect contributions, which is smaller than the $\sigma_s^{eff}$ value for our 5 nm thick WTe$_x$ film ($0.78 \times 10^5 (\hbar/2e)\ \Omega^{-1} m^{-1}$). As in the context of SHE, $\sigma_s^{eff}$ in dirty metal films is much smaller than those in the intrinsic metal regime, hence intrinsic SHE alone cannot account for the large $\sigma_s^{eff}$ value in our WTe$_x$ films. As side-jump contribution to spin Hall effect is proportional to impurity concentration and intrinsic contribution decreases drastically in the dirty metal regime, [37, 47] side-jump might play a significant role in our sputtered disordered WTe$_x$ films.

Meanwhile, as pointed out in theory, Fermi arcs surface states that connect the two Weyl points in WSM lead to large current-induced spin polarization that is one order of magnitude larger than that in Rashba systems and surface states of TIs.[48] Nevertheless, previous research has shown that the spin polarization from these surface states should be in parallel to the current injection direction, [16] which is not the case in our experiment. It is also possible that Rashba-type spin splitting observed in WTe$_2$ experimentally [49] contributes to the SOT. However, as suggested by previous work, the Rashba 2D interface should lead to strong field-like torque in thin WTe$_2$ devices, [16] while we observe minimal field-like torque in the 5 nm WTe$_x$ film (Supplementary Note S10).

**SOT and Magnetization Switching in Annealed WTe$_x$/Mo/CoFeB Heterostructures**

To demonstrate SOT-driven magnetization switching, we synthesized heterostructures composed of 5 nm thick WTe$_x$ films with low resistivity and large $\sigma_s$, which is better matched with metallic CoFeB and has higher write efficiency. Integration of sputtered WTe$_x$ films into CMOS backend processes calls for a stack design that sustains SOT properties after thermal annealing processes. Here, we insert 1-2 nm thick Mo between WTe$_x$ and CoFeB due to the excellent thermal annealing stability, while the low $\xi_{ST}$ and large spin diffusion length of Mo will not significantly alter the SOT properties of the heterostructure. [50, 51] We confirm good magnetic properties of CoFeB and minimal degradation of the WTe$_x$ film's Raman spectrum after insertion of Mo and after 300 ºC annealing (see Figure S8).

Then we utilize second-harmonic Hall measurements to quantify the damping-like SOT in annealed $WTe_x(5)/Mo(1)/CoFeB(1)$ heterostructure. Following established analysis methodology, [52] we determine there are minimal field-like torque, Oersted field, and thermoelectric effects (including anomalous Nernst and spin Seebeck effects) as discussed in Supplementary Note S10. Then, by fitting $R_{2\omega}^{xy}$ dependence on external magnetic field $H$ as shown in Figure 5**a**, damping-like effective field $H_{DL}$ can be extracted for different AC current amplitudes $I$. As shown in Figure 5**b**, we obtain $\xi_{ST}^{WTe_x} = 0.426 \pm 0.004$ through a linear fitting using equation $\xi_{ST} = \frac{2eM_S t_{CoFeB}}{\hbar} \frac{H_{DL}}{J_{WTe_x}}$, where $M_S$ is saturation magnetization and $J_{WTe_x}$ is the current density flowing through the $WTe_x$ layer, which is calculated using a parallel resistor model based on the resistivity of the $WTe_x$, Mo, and CoFeB layer: $\rho_{WTe_x} = 435\ \mu\Omega\ cm, \rho_{Mo} = 104\ \mu\Omega\ cm, \rho_{CoFeB} = 135\ \mu\Omega\ cm$ (Supplementary Note S11). We further carried out the same measurements on an in-plane Mo(6)/CoFeB(1)/MgO(2)/Ta(2) sample confirming a small $\xi_{ST}^{Mo} = 0.046 \pm 0.002$, which is at least one order of magnitude lower than that of $WTe_x$ as shown in Figure 5**b**. We find a large $\sigma_s^{eff}$ of $0.98 \times 10^5 (\hbar/2e)\ \Omega^{-1}m^{-1}$ in this annealed $WTe_x(5)/Mo(1)/CoFeB(1)$ heterostructure, which is slightly higher than that of $0.78 \times 10^5 (\hbar/2e)\ \Omega^{-1}m^{-1}$ in as deposited $WTe_x(5)/CoFeB(4)$ bilayer discussed in the last section. This confirms that large SOT is sustained after insertion of Mo layer and thermal annealing.

We next characterize the unidirectional spin Hall magnetoresistance (USMR) effect which can distinguish magnetization along *y* and –*y* directions via second-harmonic longitudinal resistance ($R_{2\omega}^{xx}$) measurements [53]. We use 300 °C-annealed $WTe_x(5)/Mo(2)/CoFeB(1)$ heterostructures here for magnetization switching experiments as the CoFeB layer in the $WTe_x(5)/Mo(1)/CoFeB(1)$ heterostructure has a negligible in-plane coercivity (see Supplementary Note S12). As shown in Figure 5**c**, when the magnetization aligns along different transverse orientations (*y*-axis), $R_{2\omega}^{xx}$ signal switches with a coercivity around 1-2 Oe. The $R_{2\omega}^{xx}$ signal vanishes at a large $H_y$ field, as shown in Figure 5**c** inset. There are two origins of USMR effects reported, namely spin-dependent electron scattering between SOT spin current and magnons in the ferromagnet [54, 55] and spin-dependent electron scattering between SOT spin currents and the magnetization [53, 56]. Because magnon is attenuated while the magnetization remains constant at large $H_y$ field, we attribute the observed field dependent USMR signal to the former origin.[57] We also determine the thermoelectric contribution to the USMR effect to be negligible (Supplementary Note S13). Following reference [54], we use total USMR per current density per total longitudinal resistance $\Delta R_{2\omega}^{xx}/JR_{xx}$ to benchmark the strength of USMR across various material stacks and device geometries. Here, $\Delta R_{2\omega}^{xx}$ is defined as the maximal change of USMR when magnetization points to –y or y direction. Notably, the $\Delta R_{2\omega}^{xx}/J_{WTe_x}R_{xx}$ value of 82.2 $ppm$ $MA^{-1}cm^2$ in this work is about 5-20 times higher than that in existing FM/HM or FM/TI bilayers (see Table S1). Note that a recent work also shows large low-temperature USMR values in TI/ferromagnetic semiconductor heterostructures originating from spin-dependent electrons scattering with the magnons.[58] We postulate that the large USMR strength in this work results from the intricate interplay between Weyl fermion-like electrons and magnons in the ferromagnet.

We then demonstrate room temperature pulsed current-driven switching of in-plane magnetization detected via the USMR effect. First a square-shaped current pulse with a pulse width of $t_{pulse}$ flows through the current channel, and subsequently the $R_{2\omega}^{xx}$ was measured under an AC read current. Similar magnetization switching curves at zero external magnetic field with opposite magnetization initialization directions as shown in Figure **5d** confirms that current-induced SOT drives the switching. The damping-like SOT from the WTe$_x$ layer has a positive sign which is consistent with exfoliated WTe$_2$ [14] and the ST-FMR results in Figure 4**b**, while field-like and Oersted field assists the SOT-driven switching (Supplementary Note S13). As the damping-like SOT from the WTe$_x$ layer has a positive sign which is opposite from that of W, the large positive SOT cannot be ascribed to the W-rich regions which is adjacent to the CoFeB layer and overlapping with the Mo region as shown in Figure S9b. The SOT efficiency from the WTe$_x$ layer might be even larger after considering a negative-sign SOT contribution from the diffused W-rich regions.

The analog and gradual switching process indicates that multi-domains form during the switching process due to the small coercivity (< 2 Oe) of the CoFeB layer [59]. While the asymmetric switching current can be attributed to a small remnant field around 0.5 Oe during measurements, as seen from the shift of maximum and minimum $R_{2\omega}^{xx}$ values away from zero field in Figure 5**c**. Note that 30.1% of the total current flows through the WTe$_x$ layer based on a parallel resistor model. (see Figure S11) The lowest switching current density $J_{WTe_x}$ achieved is

0.97 and 2.05 MA/cm$^2$ using a 100 ms pulse width. We also confirm the validity of USMR detection of in-plane magnetization switching on a Pt/CoFeB control sample. (see Figure S14)

**Discussion**

We now compare the charge-to-spin conversion properties in sputtered WTe$_x$ films with single crystalline WTe$_2$ flakes and other SOT materials. We do not discover any out-of-plane damping-like torque arising from the low-symmetry of WTe$_2$ crystal structure from both ST-FMR and second harmonic Hall measurement data (Supplementary Note S15). Presumably, this is because the out-of-plane SOTs in our nm-scale WTe$_x$ clusters, if any, when aggregated randomly in the amorphous structure, vanishes to zero on average. The $\sigma_s^{eff}$ of $0.98 \times 10^5 (\hbar/2e)\ \Omega^{-1}m^{-1}$ and $\xi_{ST}$ of 0.4 in the annealed 5 nm thick WTe$_x$ films are comparable with that of $0.88 \times 10^5 (\hbar/2e)\ \Omega^{-1}m^{-1}$ and 0.5 in exfoliated 120 nm thick WTe$_2$ flakes.[14] Our $\sigma_s^{eff}$ results are also comparable with those in the range of 0.4 – 2 [$\times 10^5 (\hbar/2e)\ \Omega^{-1}m^{-1}$] in MBE-grown Bi$_2$Se$_3$ and sputtered Bi$_x$Se$_{1-x}$, though the latter have resistivity values (1755 – 13000 $\mu\Omega\ cm$) much higher than the sputtered WTe$_x$ (435 $\mu\Omega\ cm$).[5, 19] Meanwhile, our benchmark study based on a cell-level model of in-plane SOT-MRAM suggests that the write current and energy of SOT-MRAM cell using the annealed 5 nm thick WTe$_x$ are much smaller than that of WTe$_2$, and are comparable with other sputtered FMs and TIs.[9] Note that the current and energy performance of the annealed 5 nm thick WTe$_x$ with resistivity of 435 $\mu\Omega\ cm$ is comparable with the first published work on sputtered $\beta$-W with a $\sigma_s^{eff}$ of $1.15 \times 10^5 (\hbar/2e)\ \Omega^{-1}m^{-1}$ and resistivity of 260 $\mu\Omega\ cm$,[60] but much less favorable compared with a recent work using optimized $\beta$-W with a $\sigma_s^{eff}$ of $2.6 \times 10^5 (\hbar/2e)\ \Omega^{-1}m^{-1}$ with resistivity of 238 $\mu\Omega\ cm$.[61] This motivates future research towards sputtering WTe$_x$ materials with better SOT properties, for example by tuning the WTe$_x$ stoichiometry and disorder levels. [62]In contrast to the traditional method of detecting in-plane magnetization direction using a MTJ with two in-plane magnetic layers sandwiching a tunnel barrier, the USMR effect used in this work can detect in-plane magnetization switching in a simple planar SOT layer/FM bilayer geometry.[53] The 5-20 times larger USMR strength in our heterostructure not only shows promise of USMR-based novel bilayer spintronics devices, but

also calls for more studies into the role of magnon scattering in charge-to-spin conversion physics.[54]

Last, our results show that sputtered conductive WTe$_x$ films with large and robust SOT properties are readily compatible with industry production requirements. We believe this will pave the way for future spintronics research based on sputtered two-dimensional materials [63] and topological semimetals. We also hope to stimulate future condensed matter theory and experimental studies to go beyond granular and amorphous TIs [19, 21, 25], and investigate the new class of topological disordered semimetals [26].

## Experimental Procedures

### Resource Availability

*Lead Contact*: Correspondence and requests for materials and data should be addressed to Lead Contact Xiang Li (xiangshaunli@gmail.com).

*Materials Availability*: This study did not generate new unique reagents.

*Data and Code Availability*: Supplementary information is available in the online version of the paper.

### Material deposition

The WTe$_x$ and Mo films were deposited using ion-beam sputtering techniques using a stoichiometric WTe$_2$ target and Mo target. The ion-beam sputtering was conducted using Xe gas under 0.1 mTorr at room temperature. The heavier atomic weight of Xe over Ar enables lower sputtering pressure thus larger mean free path of the sputtered target atoms. The growth rate of WTe$_2$ is around 0.29 Å/s and that of Mo is around 0.23 Å/s. Meanwhile, MgO, Ta, and Co$_{20}$Fe$_{60}$B$_{20}$ magnetron sputtering guns are also integrated into the same vacuum chamber with a base pressure below $6 \times 10^{-8}$ Torr. Hence, we can achieve *in-situ* growth of the whole stack discussed in the main text without breaking vacuum. The magnetron sputtering was conducted using Ar gas under 0.7 – 2.3 mTorr. The Ta and Co$_{20}$Fe$_{60}$B$_{20}$ layers were deposited using DC sputtering, while the MgO layers were deposited using RF sputtering from an insulating MgO target. All stacks were sputtered on thermally oxidized Si substrate. The stacks with Mo insertion were annealed at 300 °C for 30 minutes using an All-Win Rapid Thermal Process (RTP) system with Ar ambient.

**Device fabrication**

Temperature-dependent magnetoresistance and resistance data were gathered from Hall bars patterned on MgO(2)/WTe$_x$(5)/MgO(2)/Ta(2) and WTe$_x$(58)/Ta(2) stacks using a four-point probe method. The WTe$_x$/CoFeB(4.4)/MgO(2)/Ta(2) (number in parenthesis is in nm) stacks for ST-FMR measurements were fabricated into $10\ \mu m \times 40\ \mu m$ microstrips using standard photolithography and Ar ion mill techniques. The MgO(2)/WTe$_x$(5)/Mo(1 or 2)/CoFeB(1)/MgO(2)/Ta(2) (number in parenthesis is in nm) stacks for second harmonic measurements were fabricated into $10\ \mu m \times 130\ \mu m$ Hall bars using standard photolithography and Ar ion mill techniques. The patterned devices are subsequently covered with Ti(5 nm)/Au(120 nm) as contacts using photolithography and liftoff techniques.

**Film characterization**

*Raman spectroscopy:* The Raman spectrum was gathered using a Horiba Labram HR Evolution Raman System with a laser of 532 nm wavelength, a grating of 600 l/mm, and objective magnification of 100x. Each spectrum is an average of 10 captures each collected over 10 s. The laser spot size is around $0.3\ \mu m$.

*X-Ray Photoelectron Spectroscopy (XPS):* The XPS spectrum was gathered using a PHI VersaProbe System with Al (Ka) radiation (1486 eV). The focused ion gun used for *in situ* depth profiling has a sputter rate of around 2 nm/min with a beam energy of 500V 0.5 $\mu A$, and a raster size of $1\ \mu m \times 1\ \mu m$.

*Transmission Electron Microscopy (TEM):* Cross-sectional TEM samples were prepared with Focused Ion Beam technique (Thermo-Fisher Helios 460G4). At the final sample thinning stages, Ga ion beam energy was reduced from 30kV to 8kV, 2kV, and then 500V to mitigate ion damages over TEM lamella surfaces. TEM lamellas were imaged with Thermo-Fisher Metrios TEM equipped with probe Cs-correctors operated at 200kV. STEM high-angle annular dark-field (HAADF) imaging is the primary imaging mode used in this study, and the resolution is better than 0.13nm.

*Energy Dispersive X-ray Spectroscopy (EDS):* EDS experiment was carried out in STEM mode

with the probe current set to about 600pA. The EDS detector system is of SDD type with a nominal collection solid angle of 0.7 or 0.9 radians (Super-X™ or Dual-X™, respectively). EDS data analysis was performed in Bruker's Esprit EDS software. Quantification of W:Te ratio in $WTe_x$ film was obtained by using the built-in Cliff Lorimer factors of the Esprit EDS software. W signal maps (intensity of W M-lines) were processed by Principle Component Analysis (PCA) to isolate W-M lines from Si K-line and Ta-M lines.

**Device Electrical Measurements**

*Spin-Torque Ferromagnetic Resonance (ST-FMR):* ST-FMR measurements were performed on $WTe_x$/CoFeB microstrips with sizes of 60 x 40 $\mu m^2$ using the setup. The external $H$ field is oriented with a $\phi$ angle of 45º with respect to the microstrip thus current flow direction. A GHz signal was generated by an HP 83640B microwave source and was sent through the $WTe_x$/CoFeB bilayer through a T-Bias and ground-signal-ground coplanar waveguide.

*Second harmonic and pulsed switching measurements:* The second harmonic setup consists of a Keithley 6221 current source providing AC current with a frequency of 1.333 kHz, and two Stanford Research SR830 Lock-in amplifiers recording the first and second harmonic signal of longitudinal or Hall resistance of the Hall bar device. The pulse current used in the switching experiments were generated by Keithley 6221 through the square wave settings. The DC device resistance was obtained using a four-point probe method with Keithley 6221 current source and Keithley 2000 voltage meter.

**Acknowledgments**

This research was supported in part by ASCENT, one of six centers in JUMP, a Semiconductor Research Corporation (SRC) program sponsored by DARPA. Part of this work was performed at the Stanford Nano Shared Facilities (SNSF)/Stanford Nanofabrication Facility (SNF), supported by the National Science Foundation under award ECCS-1542152. The Stanford authors wish to thank NSF Center for Energy Efficient Electronics Science (E3S) and TSMC for additional financial support. The authors would also like to acknowledge Dan Ralph, Jian-Ping Wang, David Goldhaber-Gordon, Carlos H. Diaz, Donkoun Lee, Chris Hinkle, Ilan Rosen, Arturas Vailionis, Marcin Walkiewicz, and Andrey Malkovskiy for fruitful discussions.

## Author contributions

X.L. conceived and designed the research with contributions from P.L., M.D., C.B., S.-J.L., W.T., Y.S., and S.X.W.. S.X.W. supervised the study. X.L. deposited the thin films, carried out Raman spectroscopy, XPS, and second harmonic measurements. X.L. and P.L. fabricated the Hall bar and microstrip devices and carried out ST-FMR measurements. X.L., P.L., and D.Y. carried out the SQUID measurements. V.H., C.-H.N., and C.-M.L. carried out TEM and EDS studies. X.L. performed data analysis with contributions from P.L., V.H., M.D., F.X., C.-H.N., S.-J.L., W.T., Y.S., and S.X.W. X.L. wrote and revised the manuscript with input and comments from all authors.

## Declaration of interests

The authors declare no competing interests.

# References

1. Hasan, M. Z.; Kane, C. L., Colloquium: Topological insulators. *Reviews of Modern Physics* **2010,** *82* (4), 3045-3067.
2. Qi, X.-L.; Zhang, S.-C., Topological insulators and superconductors. *Reviews of Modern Physics* **2011,** *83* (4), 1057-1110.
3. Hsieh, D.; Xia, Y.; Qian, D.; Wray, L.; Dil, J. H.; Meier, F.; Osterwalder, J.; Patthey, L.; Checkelsky, J. G.; Ong, N. P.; Fedorov, A. V.; Lin, H.; Bansil, A.; Grauer, D.; Hor, Y. S.; Cava, R. J.; Hasan, M. Z., A tunable topological insulator in the spin helical Dirac transport regime. *Nature* **2009,** *460*, 1101.
4. Li, C. H.; van 't Erve, O. M. J.; Robinson, J. T.; Liu, Y.; Li, L.; Jonker, B. T., Electrical detection of charge-current-induced spin polarization due to spin-momentum locking in Bi2Se3. *Nature Nanotechnology* **2014,** *9*, 218.
5. Mellnik, A. R.; Lee, J. S.; Richardella, A.; Grab, J. L.; Mintun, P. J.; Fischer, M. H.; Vaezi, A.; Manchon, A.; Kim, E. A.; Samarth, N.; Ralph, D. C., Spin-transfer torque generated by a topological insulator. *Nature* **2014,** *511* (7510), 449-51.
6. Wang, Y.; Zhu, D. P.; Wu, Y.; Yang, Y. M.; Yu, J. W.; Ramaswamy, R.; Mishra, R.; Shi, S. Y.; Elyasi, M.; Teo, K. L.; Wu, Y. H.; Yang, H., Room temperature magnetization switching in topological insulator-ferromagnet heterostructures by spin-orbit torques. *Nature Communications* **2017,** *8*.
7. Wang, H.; Kally, J.; Lee, J. S.; Liu, T.; Chang, H.; Hickey, D. R.; Mkhoyan, K. A.; Wu, M.; Richardella, A.; Samarth, N., Surface-State-Dominated Spin-Charge Current Conversion in Topological-Insulator--Ferromagnetic-Insulator Heterostructures. *Physical Review Letters* **2016,** *117* (7), 076601.
8. Liu, M.; Zhang, J.; Chang, C.-Z.; Zhang, Z.; Feng, X.; Li, K.; He, K.; Wang, L.-l.; Chen, X.; Dai, X.; Fang, Z.; Xue, Q.-K.; Ma, X.; Wang, Y., Crossover between Weak Antilocalization and Weak Localization in a Magnetically Doped Topological Insulator. *Physical Review Letters* **2012,** *108* (3), 036805.
9. Li, X.; Lin, S.; Dc, M.; Liao, Y.; Yao, C.; Naeemi, A.; Tsai, W.; Wang, S. X., Materials Requirements of High-Speed and Low-Power Spin-Orbit-Torque Magnetic Random-Access Memory. *IEEE Journal of the Electron Devices Society* **2020,** *8*, 674-680.
10. Feng, B.; Chan, Y.-H.; Feng, Y.; Liu, R.-Y.; Chou, M.-Y.; Kuroda, K.; Yaji, K.; Harasawa, A.; Moras, P.; Barinov, A.; Malaeb, W.; Bareille, C.; Kondo, T.; Shin, S.; Komori, F.; Chiang, T.-C.; Shi, Y.; Matsuda, I., Spin texture in type-II Weyl semimetal WTe2. *Physical Review B* **2016,** *94* (19).
11. Zyuzin, A. A.; Wu, S.; Burkov, A. A., Weyl semimetal with broken time reversal and inversion symmetries. *Physical Review B* **2012,** *85* (16), 165110.
12. Soluyanov, A. A.; Gresch, D.; Wang, Z.; Wu, Q.; Troyer, M.; Dai, X.; Bernevig, B. A., Type-II Weyl semimetals. *Nature* **2015,** *527*, 495.
13. Ali, M. N.; Xiong, J.; Flynn, S.; Tao, J.; Gibson, Q. D.; Schoop, L. M.; Liang, T.; Haldolaarachchige, N.; Hirschberger, M.; Ong, N. P.; Cava, R. J., Large, non-saturating magnetoresistance in WTe2. *Nature* **2014,** *514* (7521), 205-+.
14. Shi, S.; Liang, S.; Zhu, Z.; Cai, K.; Pollard, S. D.; Wang, Y.; Wang, J.; Wang, Q.; He, P.; Yu, J.; Eda, G.; Liang, G.; Yang, H., All-electric magnetization switching and Dzyaloshinskii-Moriya interaction in WTe2/ferromagnet heterostructures. *Nat Nanotechnol* **2019,** *14* (10), 945-949.
15. Zhao, B.; Khokhriakov, D.; Zhang, Y.; Fu, H.; Karpiak, B.; Hoque, A. M.; Xu, X.; Jiang, Y.; Yan, B.; Dash, S. P., Observation of charge to spin conversion in Weyl semimetal ${\mathrm{WTe}}_{2}$ at room temperature. *Physical Review Research* **2020,** *2* (1), 013286.
16. Li, P.; Wu, W. K.; Wen, Y.; Zhang, C. H.; Zhang, J. W.; Zhang, S. F.; Yu, Z. M.; Yang, S. Y. A.; Manchon, A.; Zhang, X. X., Spin-momentum locking and spin-orbit torques in magnetic nano-heterojunctions composed of Weyl semimetal WTe2. *Nature Communications* **2018,** *9*.

17. MacNeill, D.; Stiehl, G. M.; Guimaraes, M. H. D.; Buhrman, R. A.; Park, J.; Ralph, D. C., Control of spin–orbit torques through crystal symmetry in WTe2/ferromagnet bilayers. *Nature Physics* **2016,** *13*, 300.
18. Fan, Y.; Upadhyaya, P.; Kou, X.; Lang, M.; Takei, S.; Wang, Z.; Tang, J.; He, L.; Chang, L. T.; Montazeri, M.; Yu, G.; Jiang, W.; Nie, T.; Schwartz, R. N.; Tserkovnyak, Y.; Wang, K. L., Magnetization switching through giant spin-orbit torque in a magnetically doped topological insulator heterostructure. *Nat Mater* **2014,** *13* (7), 699-704.
19. Dc, M.; Grassi, R.; Chen, J. Y.; Jamali, M.; Reifsnyder Hickey, D.; Zhang, D.; Zhao, Z.; Li, H.; Quarterman, P.; Lv, Y.; Li, M.; Manchon, A.; Mkhoyan, K. A.; Low, T.; Wang, J. P., Room-temperature high spin-orbit torque due to quantum confinement in sputtered BixSe(1-x) films. *Nat Mater* **2018,** *17* (9), 800-807.
20. Banerjee, A.; Deb, O.; Majhi, K.; Ganesan, R.; Sen, D.; Anil Kumar, P. S., Granular topological insulators. *Nanoscale* **2017,** *9* (20), 6755-6764.
21. Corbae, P.; Ciocys, S.; Varjas, D.; Zeltmann, S.; Stansbury, C. H.; Molina-Ruiz, M.; Chen, Z.; Wang, L.-W.; Minor, A. M.; Grushin, A. G., Evidence for topological surface states in amorphous Bi $_{2}$ Se $_{3}$. *arXiv preprint arXiv:1910.13412* **2019**.
22. Weaire, D.; Thorpe, M. F., Electronic Properties of an Amorphous Solid. I. A Simple Tight-Binding Theory. *Physical Review B* **1971,** *4* (8), 2508-2520.
23. Haydock, R.; Heine, V.; Kelly, M. J., Electronic structure based on the local atomic environment for tight-binding bands. II. *Journal of Physics C: Solid State Physics* **1975,** *8* (16), 2591-2605.
24. Mitchell, N. P.; Nash, L. M.; Hexner, D.; Turner, A. M.; Irvine, W. T. M., Amorphous topological insulators constructed from random point sets. *Nature Physics* **2018,** *14* (4), 380-385.
25. Agarwala, A.; Shenoy, V. B., Topological Insulators in Amorphous Systems. *Physical Review Letters* **2017,** *118* (23), 236402.
26. Yang, Y.-B.; Qin, T.; Deng, D.-L.; Duan, L. M.; Xu, Y., Topological Amorphous Metals. *Physical Review Letters* **2019,** *123* (7), 076401.
27. Jiang, Y. C.; Gao, J.; Wang, L., Raman fingerprint for semi-metal WTe2 evolving from bulk to monolayer. *Sci Rep-Uk* **2016,** *6*, 19624.
28. Chen, K.; Chen, Z.; Wan, X.; Zheng, Z.; Xie, F.; Chen, W.; Gui, X.; Chen, H.; Xie, W.; Xu, J., A Simple Method for Synthesis of High-Quality Millimeter-Scale 1T′ Transition-Metal Telluride and Near-Field Nanooptical Properties. *Advanced Materials* **2017,** *29* (38), 1700704.
29. Hikami, S.; Larkin, A. I.; Nagaoka, Y., Spin-Orbit Interaction and Magnetoresistance in the Two Dimensional Random System. *Progress of Theoretical Physics* **1980,** *63* (2), 707-710.
30. Lee, P. A.; Ramakrishnan, T. V., Disordered electronic systems. *Reviews of Modern Physics* **1985,** *57* (2), 287-337.
31. Zhao, B.; Cheng, P.; Pan, H.; Zhang, S.; Wang, B.; Wang, G.; Xiu, F.; Song, F., Weak antilocalization in Cd3As2 thin films. *Sci Rep-Uk* **2016,** *6*, 22377.
32. Schumann, T.; Galletti, L.; Kealhofer, D. A.; Kim, H.; Goyal, M.; Stemmer, S., Observation of the Quantum Hall Effect in Confined Films of the Three-Dimensional Dirac Semimetal ${\mathrm{Cd}}_{3}{\mathrm{As}}_{2}$. *Physical Review Letters* **2018,** *120* (1), 016801.
33. Lu, H.-Z.; Shen, S.-Q., Weak antilocalization and localization in disordered and interacting Weyl semimetals. *Physical Review B* **2015,** *92* (3), 035203.
34. Dai, X.; Lu, H.-Z.; Shen, S.-Q.; Yao, H., Detecting monopole charge in Weyl semimetals via quantum interference transport. *Physical Review B* **2016,** *93* (16), 161110.
35. Wang, L.; Gutiérrez-Lezama, I.; Barreteau, C.; Ubrig, N.; Giannini, E.; Morpurgo, A. F., Tuning magnetotransport in a compensated semimetal at the atomic scale. *Nature Communications* **2015,** *6*, 8892.

36. Mott, N. F., Conduction in non-crystalline materials. *The Philosophical Magazine: A Journal of Theoretical Experimental and Applied Physics* **1969,** *19* (160), 835-852.
37. Vignale, G., Ten Years of Spin Hall Effect. *J Supercond Nov Magn* **2009,** *23* (1), 3.
38. Nagaosa, N.; Sinova, J.; Onoda, S.; MacDonald, A. H.; Ong, N. P., Anomalous Hall effect. *Reviews of Modern Physics* **2010,** *82* (2), 1539-1592.
39. Checkelsky, J. G.; Hor, Y. S.; Cava, R. J.; Ong, N. P., Bulk Band Gap and Surface State Conduction Observed in Voltage-Tuned Crystals of the Topological Insulator ${\mathrm{Bi}}_{2}{\mathrm{Se}}_{3}$. *Physical Review Letters* **2011,** *106* (19), 196801.
40. Ren, Z.; Taskin, A. A.; Sasaki, S.; Segawa, K.; Ando, Y., Large bulk resistivity and surface quantum oscillations in the topological insulator ${\text{Bi}}_{2}{\text{Te}}_{2}\text{Se}$. *Physical Review B* **2010,** *82* (24), 241306.
41. Du, R.; Hsu, H.-C.; Balram, A. C.; Yin, Y.; Dong, S.; Dai, W.; Zhao, W.; Kim, D.; Yu, S.-Y.; Wang, J.; Li, X.; Mohney, S. E.; Tadigadapa, S.; Samarth, N.; Chan, M. H. W.; Jain, J. K.; Liu, C.-X.; Li, Q., Robustness of topological surface states against strong disorder observed in $\mathrm{B}{\mathrm{i}}_{2}\mathrm{T}{\mathrm{e}}_{3}$ nanotubes. *Physical Review B* **2016,** *93* (19), 195402.
42. Liao, J.; Ou, Y.; Feng, X.; Yang, S.; Lin, C.; Yang, W.; Wu, K.; He, K.; Ma, X.; Xue, Q.-K.; Li, Y., Observation of Anderson Localization in Ultrathin Films of Three-Dimensional Topological Insulators. *Physical Review Letters* **2015,** *114* (21), 216601.
43. Liu, W. L.; Chen, M. L.; Li, X. X.; Dubey, S.; Xiong, T.; Dai, Z. M.; Yin, J.; Guo, W. L.; Ma, J. L.; Chen, Y. N.; Tan, J.; Li, D.; Wang, Z. H.; Li, W.; Bouchiat, V.; Sun, D. M.; Han, Z.; Zhang, Z. D., Effect of aging-induced disorder on the quantum transport properties of few-layer WTe2. *2D Materials* **2016,** *4* (1), 011011.
44. Mleczko, M. J.; Xu, R. L.; Okabe, K.; Kuo, H.-H.; Fisher, I. R.; Wong, H. S. P.; Nishi, Y.; Pop, E., High Current Density and Low Thermal Conductivity of Atomically Thin Semimetallic WTe2. *ACS Nano* **2016,** *10* (8), 7507-7514.
45. Liu, L.; Lee, O. J.; Gudmundsen, T. J.; Ralph, D. C.; Buhrman, R. A., Current-Induced Switching of Perpendicularly Magnetized Magnetic Layers Using Spin Torque from the Spin Hall Effect. *Physical Review Letters* **2012,** *109* (9), 096602.
46. Zhou, J.; Qiao, J.; Bournel, A.; Zhao, W., Intrinsic spin Hall conductivity of the semimetals ${\mathrm{MoTe}}_{2}$ and ${\mathrm{WTe}}_{2}$. *Physical Review B* **2019,** *99* (6), 060408.
47. Fert, A.; Levy, P. M., Spin Hall Effect Induced by Resonant Scattering on Impurities in Metals. *Physical Review Letters* **2011,** *106* (15), 157208.
48. Johansson, A.; Henk, J.; Mertig, I., Edelstein effect in Weyl semimetals. *Physical Review B* **2018,** *97* (8), 085417.
49. Li, Q.; Yan, J.; Yang, B.; Zang, Y.; Zhang, J.; He, K.; Wu, M.; Zhao, Y.; Mandrus, D.; Wang, J.; Xue, Q.; Chi, L.; Singh, D. J.; Pan, M., Interference evidence for Rashba-type spin splitting on a semimetallic $\mathrm{WT}{\mathrm{e}}_{2}$ surface. *Physical Review B* **2016,** *94* (11), 115419.
50. Mosendz, O.; Pearson, J. E.; Fradin, F. Y.; Bauer, G. E. W.; Bader, S. D.; Hoffmann, A., Quantifying Spin Hall Angles from Spin Pumping: Experiments and Theory. *Physical Review Letters* **2010,** *104* (4).
51. Liu, T.; Zhang, Y.; Cai, J. W.; Pan, H. Y., Thermally robust Mo/CoFeB/MgO trilayers with strong perpendicular magnetic anisotropy. *Sci Rep* **2014,** *4*, 5895.
52. Avci, C. O.; Garello, K.; Gabureac, M.; Ghosh, A.; Fuhrer, A.; Alvarado, S. F.; Gambardella, P., Interplay of spin-orbit torque and thermoelectric effects in ferromagnet/normal-metal bilayers. *Physical Review B* **2014,** *90* (22), 224427.

53. Avci, C. O.; Garello, K.; Ghosh, A.; Gabureac, M.; Alvarado, S. F.; Gambardella, P., Unidirectional spin Hall magnetoresistance in ferromagnet/normal metal bilayers. *Nature Physics* **2015,** *11*, 570.
54. Yasuda, K.; Tsukazaki, A.; Yoshimi, R.; Takahashi, K. S.; Kawasaki, M.; Tokura, Y., Large Unidirectional Magnetoresistance in a Magnetic Topological Insulator. *Physical Review Letters* **2016,** *117* (12), 127202.
55. Cheng, Y.; Chen, K.; Zhang, S., Interplay of magnon and electron currents in magnetic heterostructure. *Physical Review B* **2017,** *96* (2), 024449.
56. Zhang, S. S. L.; Vignale, G., Theory of unidirectional spin Hall magnetoresistance in heavy-metal/ferromagnetic-metal bilayers. *Physical Review B* **2016,** *94* (14), 140411.
57. Avci, C. O.; Mendil, J.; Beach, G. S. D.; Gambardella, P., Origins of the Unidirectional Spin Hall Magnetoresistance in Metallic Bilayers. *Physical Review Letters* **2018,** *121* (8), 087207.
58. Duy Khang, N. H.; Hai, P. N., Giant unidirectional spin Hall magnetoresistance in topological insulator – ferromagnetic semiconductor heterostructures. *Journal of Applied Physics* **2019,** *126* (23), 233903.
59. Yasuda, K.; Tsukazaki, A.; Yoshimi, R.; Kondou, K.; Takahashi, K. S.; Otani, Y.; Kawasaki, M.; Tokura, Y., Current-Nonlinear Hall Effect and Spin-Orbit Torque Magnetization Switching in a Magnetic Topological Insulator. *Physical Review Letters* **2017,** *119* (13).
60. Pai, C.-F.; Liu, L.; Li, Y.; Tseng, H. W.; Ralph, D. C.; Buhrman, R. A., Spin transfer torque devices utilizing the giant spin Hall effect of tungsten. *Applied Physics Letters* **2012,** *101* (12), 122404.
61. Takeuchi, Y.; Zhang, C.; Okada, A.; Sato, H.; Fukami, S.; Ohno, H., Spin-orbit torques in high-resistivity-W/CoFeB/MgO. *Applied Physics Letters* **2018,** *112* (19), 192408.
62. Huang, J. H.; Deng, K. Y.; Liu, P. S.; Wu, C. T.; Chou, C. T.; Chang, W. H.; Lee, Y. J.; Hou, T. H., Large‐Area 2D Layered MoTe2 by Physical Vapor Deposition and Solid‐Phase Crystallization in a Tellurium‐Free Atmosphere. *Advanced Materials Interfaces* **2017,** *4* (17).
63. Lin, X.; Yang, W.; Wang, K. L.; Zhao, W., Two-dimensional spintronics for low-power electronics. *Nature Electronics* **2019,** *2* (7), 274-283.

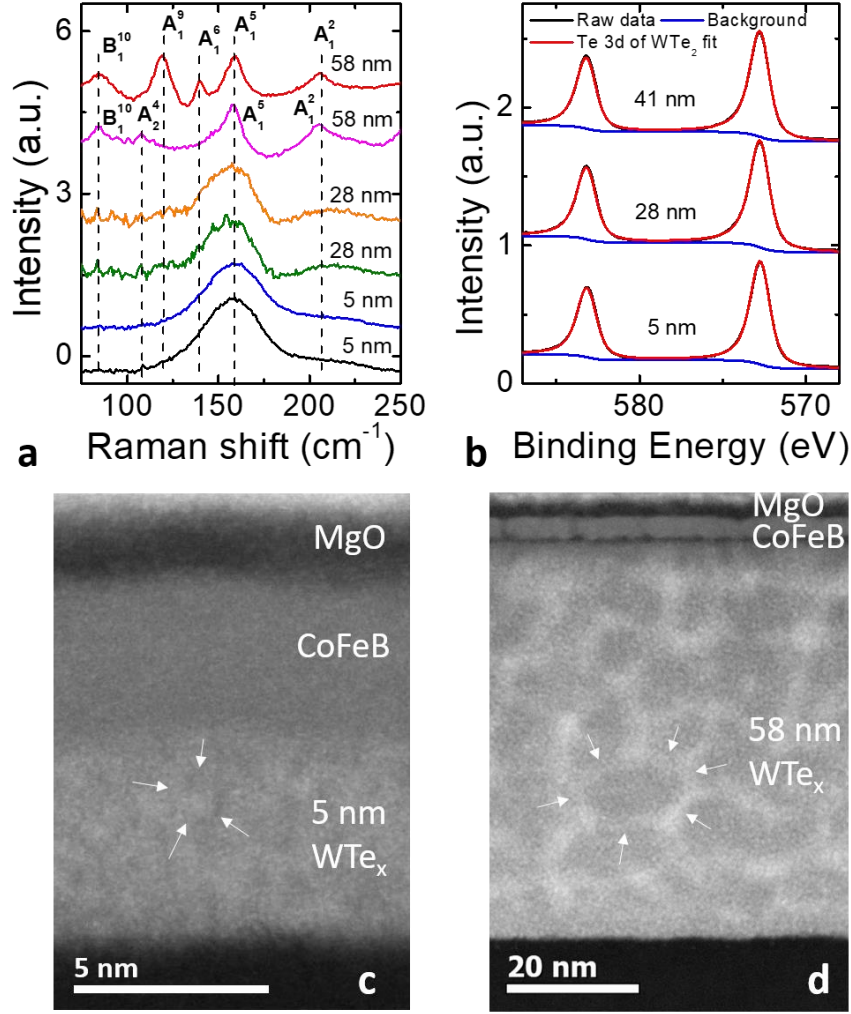

**Figure 1. Material properties of WTe$_x$ films with different thicknesses. a**, Raman spectroscopy of uncapped WTe$_x$ films with different thicknesses measured the same day when deposited. The Raman vibrational modes are labeled. **b,** X-ray Photoelectron Spectroscopy (XPS) profile of Te 3d binding energy region for WTe$_x$ films with different thicknesses, the data were obtained after *in situ* Ar ion sputtering of the capping layers on WTe$_x$ films. A doublet of Te 3d$_{5/2}$ and Te 3d$_{3/2}$ peaks are separate by 10.4 eV. **c,** Cross-section scanning transmission electron microscopy (STEM) image of WTe$_x$(5)/CoFeB(4) bilayer (hereafter, all numbers in parenthesis are in nm). **d,** Cross-section STEM image of WTe$_x$(58)/CoFeB(4) bilayer.

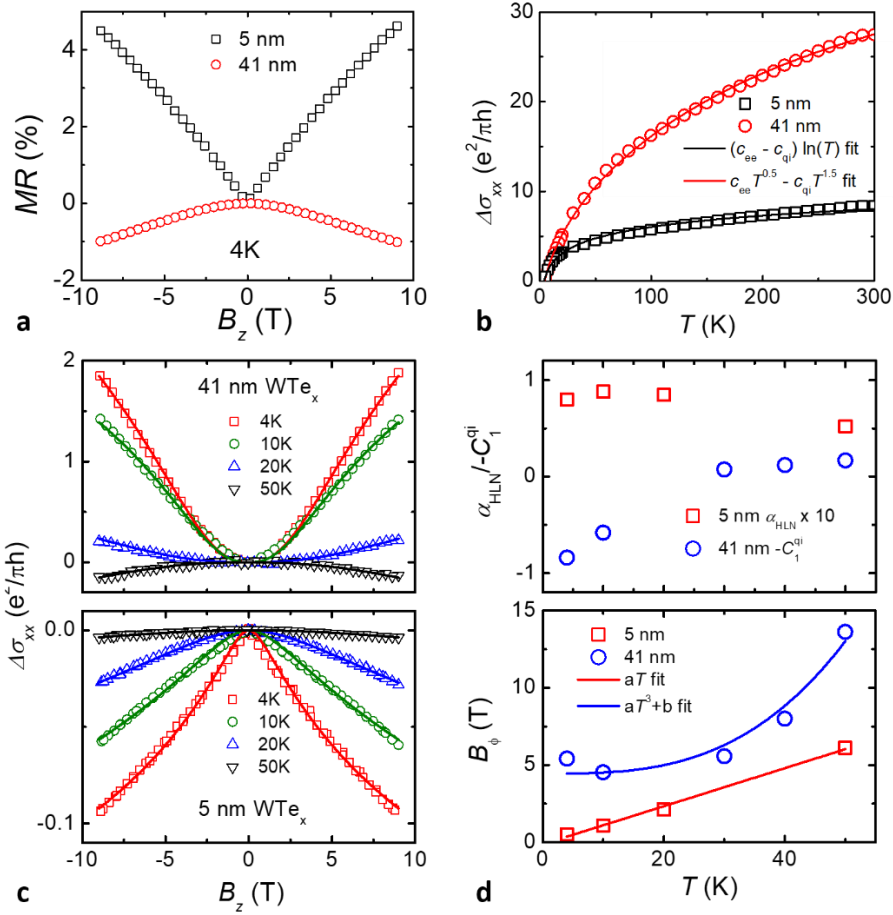

**Figure 2. Magnetoresistance and temperature-dependent resistance measurements of WTe$_x$ films with different thicknesses. a,** Magnetoresistance (MR) dependence on perpendicular magnetic field ($B_z$) for WTe$_x$ capped with MgO/Ta, measured at 4K. **b,** Change in sheet conductance ($\Delta\sigma_{xx}$) dependence on temperature. The 5 nm and 41 nm data below 150 K are fitted using $(c_{ee} - c_{qi})\ln(T)$ and $c_{ee}T^{0.5} - c_{qi}T^{1.5}$ equations, respectively. The fitted values are $c_{ee} - c_{qi} = 1.83 \pm 0.02$ for the 5 nm data, and $c_{ee} = 2.16 \pm 0.02$, $c_{qi} = (1.07 \pm 0.05) \times 10^{-3}$ for the 41 nm data, respectively. **c,** Change in $\Delta\sigma_{xx}$ dependence on perpendicular magnetic field ($B_z$) for WTe$_x$ capped with MgO/Ta, measured at 4, 10, 20, and 50 K. The 5 nm and 41 nm thick WTe$_x$ sample data are fitted based on the 2D Hikami-Larkin-Nagaoka equation [29] and 3D WL/WAL formula proposed in [34], respectively. **d,** Dependence of phase coherence characteristic field ($B_\phi$) and fit parameters ($\alpha_{HLN}, -C_1^{qi}$) on temperature. The $B_\phi$ values for 5 nm and 41 nm data are fitted using $aT$ and $aT^3 + b$ functions, respectively. Note that the fit based on Equation 2 for the 41 nm data yields $C_2^{qi}$ value of zero for all temperatures.

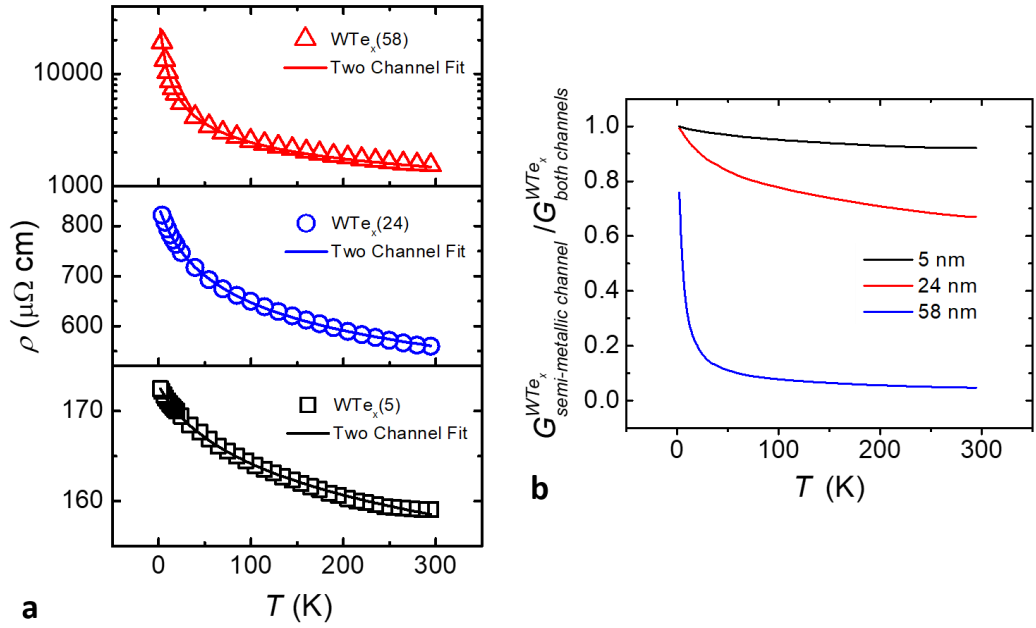

**Figure 3. Temperature-dependent resistance measurements of WTe$_x$ films with different thicknesses. a,** Temperature dependent resistivity ($\rho$) of 5 nm, 24 nm, and 58 nm WTe$_x$ extracted from WTe$_x$/CoFeB bilayer data and fit using a two-channel conduction model. **b,** Contribution of the semi-metallic channel to the total conductance as a function of temperature for WTe$_x$ with different thicknesses.

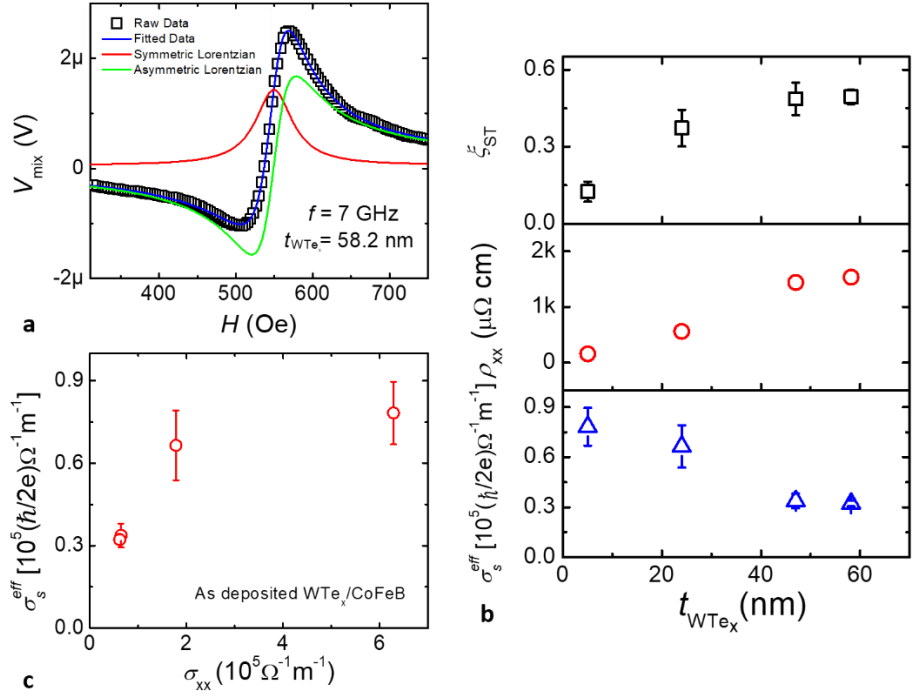

**Figure 4. Room-temperature charge-to-spin conversion in WTe$_x$/CoFeB bilayers with different WTe$_x$ thicknesses. a**, Representative mixing voltage obtained in ST-FMR measurement as a function of applied magnetic field $H$ strength for WTe$_x$ thickness of 58 nm measured at 7 GHz RF excitation. The raw data is being fitted by a sum of symmetric and asymmetric Lorentzian. **b**, $\xi_{ST}$, resistivity $\rho_{xx}$, and effective spin Hall co nductivity $\sigma_s^{eff}$ dependence on WTe$_x$ thickness from ST-FMR measurements. Each data point is averaged from ST-FMR measurements at 6 and 7 GHz. **c**, $\sigma_s^{eff}$ as a function of longitudinal conductivity $\sigma_{xx}$ for as-deposited WTe$_x$/CoFeB heterostructures. All data were measured at room temperature.

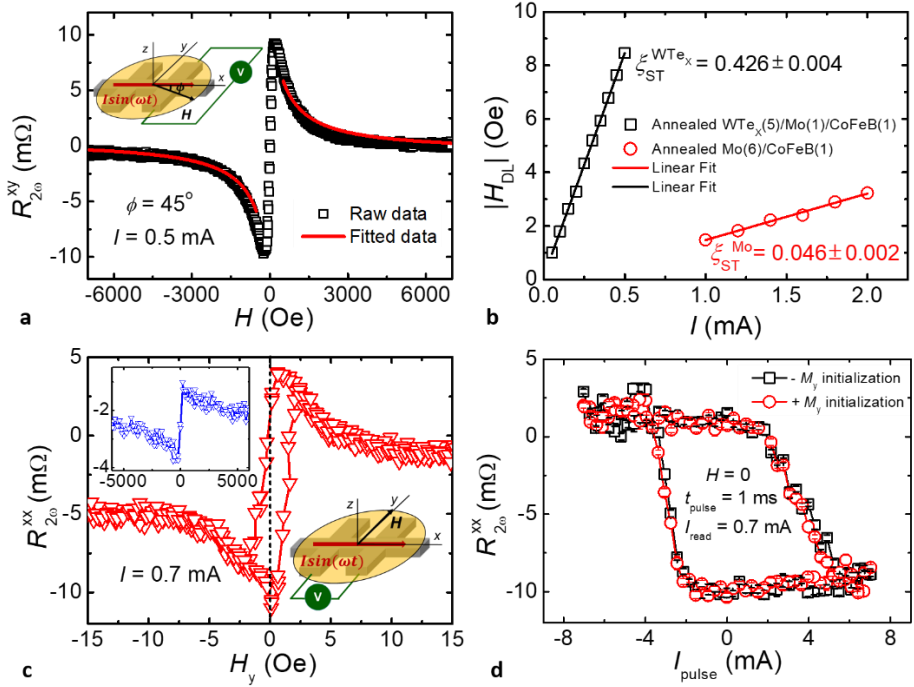

**Figure 5. Room-temperature second harmonic measurements of SOT and pulsed SOT-induced switching in WTe$_x$/Mo/CoFeB-based heterostructure. a,** Fitted in-plane $H$ dependence of $R_{2\omega}^{xy}$ graph with $\phi = 45°$ and $I = 0.5\,mA$ for WTe$_x$(5)/Mo(1)/CoFeB(1)/MgO(2)/Ta(2) heterostructure. Inset shows the experimental setup of second harmonic measurement with an input of sinusoidal AC current $I\sin(\omega t)$. **b,** Extracted absolute value of damping-like field as a function of AC current amplitude $I$ for WTe$_x$(5)/Mo(1)/CoFeB(1)/MgO(2)/Ta(2) and Mo(6)/CoFeB(1)/MgO(2)/Ta(2) heterostructure, both annealed at 300ºC for 30 minutes. **c,** Second-harmonic longitudinal resistance ($R_{2\omega}^{xx}$) as a function of the in-plane magnetic field along the y-axis ($H_y$) under an AC current amplitude of $I = 0.7\,mA$ for WTe$_x$(5)/Mo(2)/CoFeB(1)/MgO(2)/Ta(2) heterostructure annealed at 300 ºC for 30 minutes. The dotted line corresponds to $H_y = 0$. The top-left inset shows $R_{2\omega}^{xx}$ under a wider range of $H_y$ from -6000 Oe to 6000 Oe. The bottom-right inset shows the schematic of the second harmonic measurement with an input of sinusoidal current $I\sin(\omega t)$. **d,** $R_{2\omega}^{xx}$ measured as a function of pulse current amplitude $I_{pulse}$ under zero external field. The current pulse width is 1 ms. The read current amplitude is 0.7 mA. The red and black curves correspond to initialization of magnetization along $-M_y$ and $M_y$ direction respectively. All measurements were done at room temperature.

**Note S1. Raman spectrum analysis of uncapped WTe$_x$ thin films.**

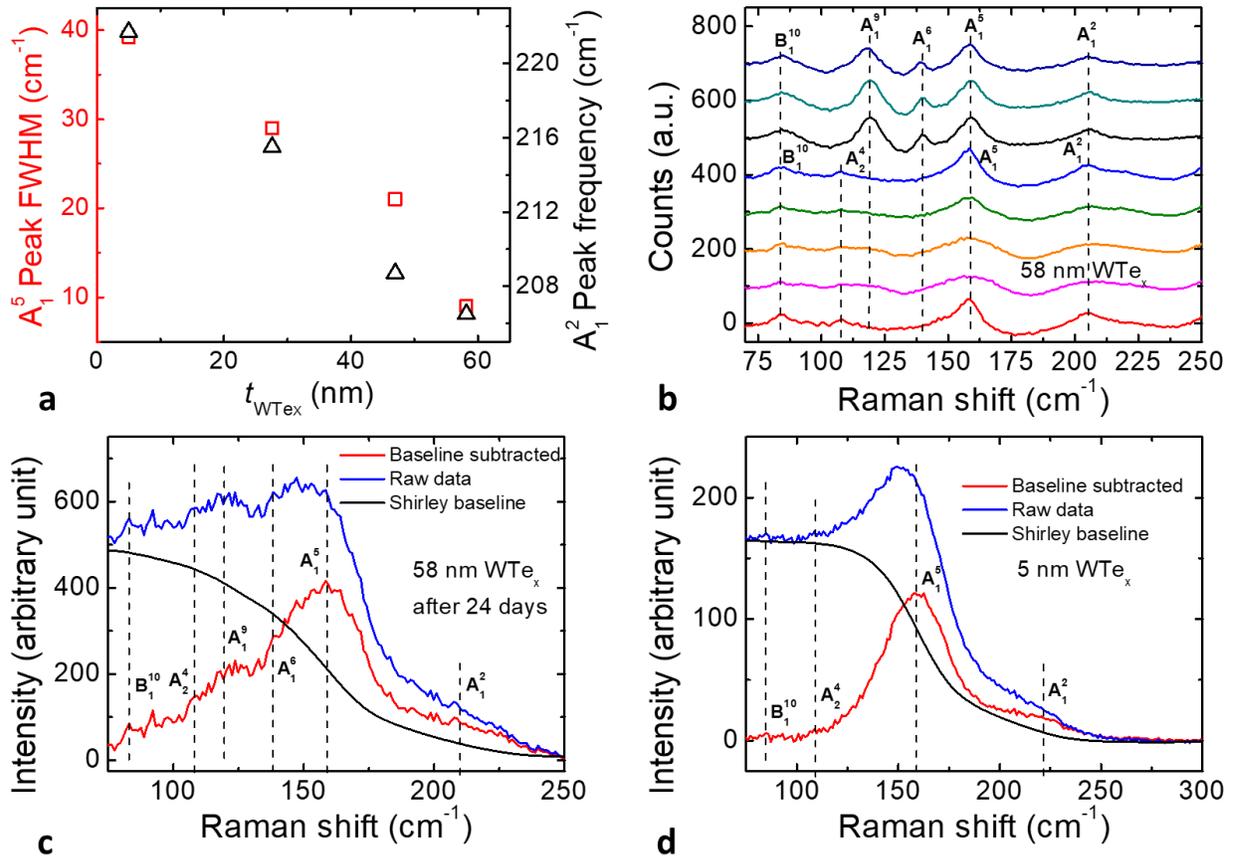

**Figure S1. Raman spectrum analysis of uncapped WTe$_x$ thin films. a,** Raman $A_1^5$ mode full width at half maximum (FWHM) and Raman $A_1^2$ mode peak frequency as a function of different WTe$_x$ thicknesses. **b,** Additional Raman spectrum of various regions measured on the same 58 nm thick WTe$_x$ sample. The film was measured the same day when deposited. **c,** Raman spectrum of 58 nm thick WTe$_x$ thin film before and after subtracting a Shirley baseline. The film was measured after exposure to air for 24 days. Compared with the Raman spectrum in **b** without any baseline, we attribute this baseline to oxidation of WTe$_x$ after exposure to air. **d,** Raman spectrum of 5 nm thick WTe$_x$ thin film before and after subtracting a Shirley baseline. Though the film was measured the same day when deposited, the much smaller thickness of 5 nm results in enhanced surface oxidation, thus a similar baseline as in **c**.

**Note S2. HR-STEM dark-field images and corresponding digital diffractograms of WTe$_x$ with different thicknesses**

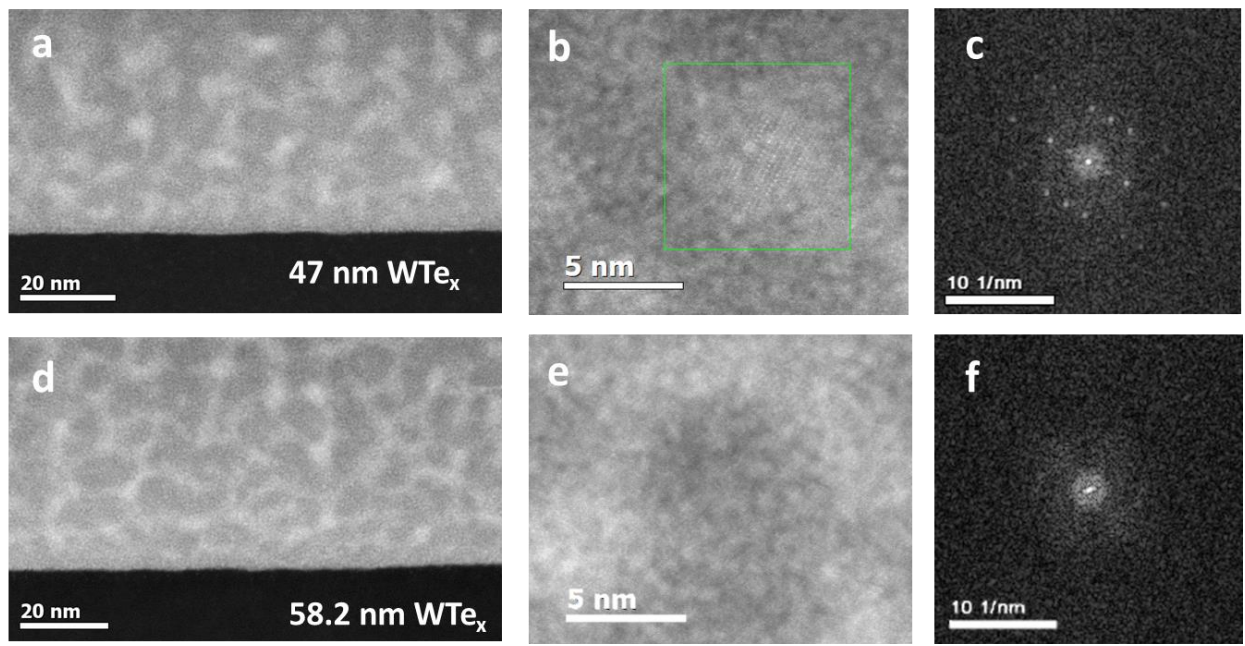

**Figure S2. HR-STEM dark-field images and corresponding digital diffractograms of WTe$_x$ with different thicknesses, with 58.2 nm for a-c, and 47 nm for d-f**. **a, b, d, e.** Cross-section STEM images of different field-of-views. A small crystallite is outlined by the green box in **b**. The digital diffractogram of the selected crystallite in **b** is shown in **c**. While **f** is the digital diffractogram of the total area as shown in **e**. All measurements were done at room temperature.

**Note S3. EDS data analysis for WTe$_x$ films with different thicknesses**

EDS data shown in Figure 3 are manually processed and various de-noise/smooth schemes are used. We have compared these manually processed data with those processed directly with EDS software as provided with TEM-EDS system (Bruker Esprit 2 software) to confirm that no artifacts are generated from the de-noise process. The detailed process of generating the W:Te atomic ratio is as follows:

First, as shown in Figure S3**d-f**, W elemental signal map is obtained by means of de-noising with PCA (Principle Component Analysis) followed by varimax matrix rotation in spatial domain. Essentially this approach will filter out noise that is not related to W signal (W M lines). The mapping result with this approach has been compared with normal EDS mapping conducted in EDS software. It can be concluded that the un-uniform nature of W "segregation" in WTe$_x$ layer is real, not data processing artifact.

Second, EDS elemental Profile (W and Te signals) in Figure S3**g-h** are obtained as follows. Original 2-dimensional (2D) EDS Spectrum Image (SI) data set is first de-noised with Gaussian smooth operation on each energy plan. Then a 1-dimensional EDS SI is extracted from the area of interest and following elemental profiles (Te and W) are then deduced from this "de-noised" 1D EDS SI. Te and W signal profiles are extracted from 1D SI from energy range of 3.6-4.4kV (Te-L lines) and 1.6-1.9kV (W-M line), respectively. Both profiles are then normalized with respect to the mean value of Te signal profiles. Te/W atomic ratio $x$ is calculated from Te/W signal ratio profiles with effective k-factor of 0.529 (i.e. Te/W atomic % ratio = Te-L/W-M signal ratio/0.529; this k-factor is derived from built-in k-factor database of Bruker's Esprit 2 Software). Again, EDS profiles obtained this way are also compared with those directly processed by Bruker Esprit EDS software. They are comparable except that profiles processed with above process exhibit less variation primarily due to the Gaussian smooth performed at each energy plan of the 2D EDS SI.

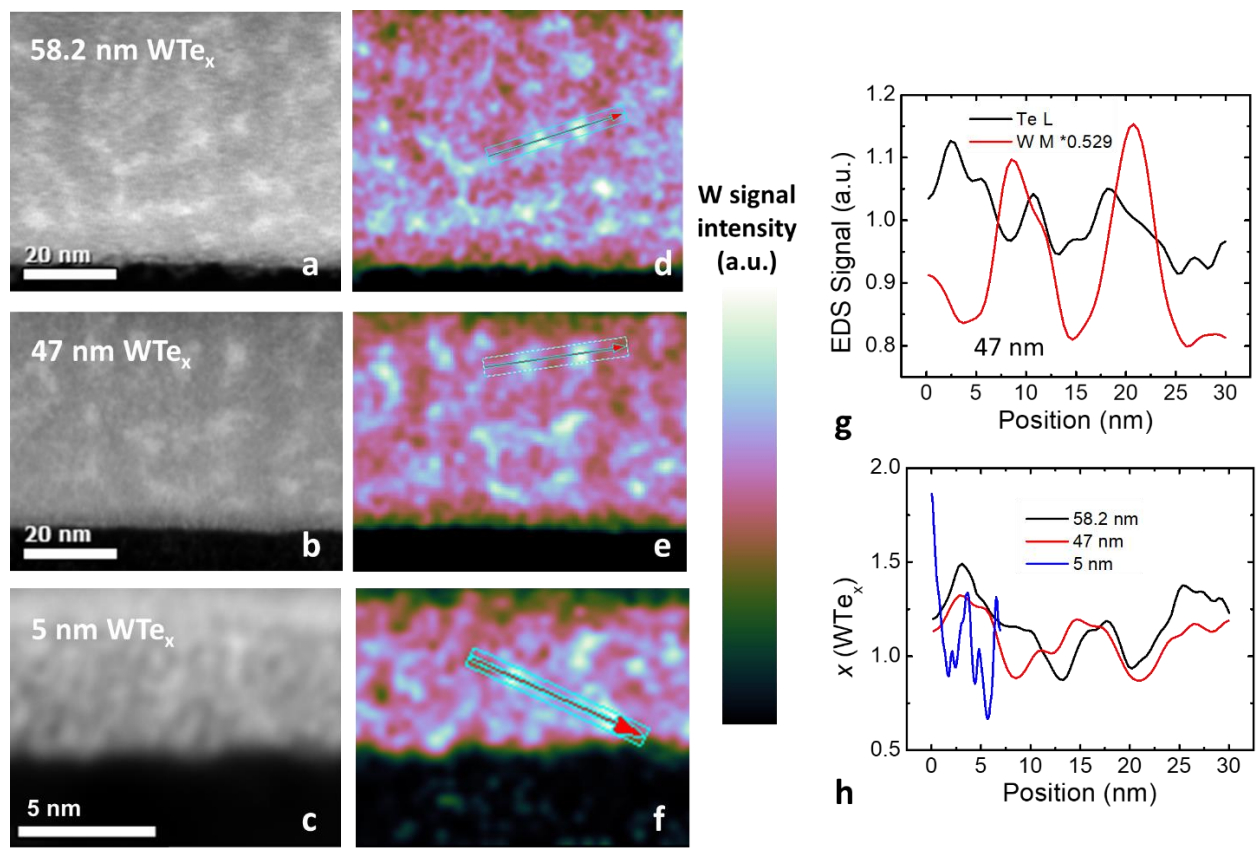

**Figure S3. STEM and EDS characterization of WTe$_x$ thin films with different thicknesses. a-c,** High-resolution STEM image, and **d-f,** W EDS intensity mapping of 58 nm, 47 nm, and 5nm WTe$_x$ respectively. **g,** Integrated W and Te EDS signal in the boxed region of WTe$_x$ (47)/CoFeB(4) sample as shown in **d**, the W signal is normalized to that of Te signal using effective k-factor 0.529. **h,** Te:W atomic ratio $x$ in the integrated boxed regions of WTe$_x$ with the three thicknesses as shown in **d-f**.

**Note S4. Comparison of magnetoresistance in WTe$_x$ thin film with published literature**

The WAL behavior for the 5 nm WTe$_x$ sample shown in Figure 2**a** has been found in thin WTe$_2$ flakes with strong disorder [1, 2] and differs from reported classical quadratic MR [1, 3] and positive linear MR in thick WTe$_2$ flakes or single crystals [4, 5]. As shown in Figure S4**a**, the $dR/dB_z$ derivative values keep decreasing at 4 K, while keep increasing at 10 and 20 K, which deviate from a linear $B_z$-dependent magnetoresistance behavior where the derivative values stay constant.

Meanwhile, as shown in Figure S4**b**, we find almost identical negative MR when the magnetic field is aligned along different directions in the 41 nm WTe$_x$ sample. This differs from chiral anomaly-induced negative MR which appears only when the magnetic field is parallel to the current.[6, 7].

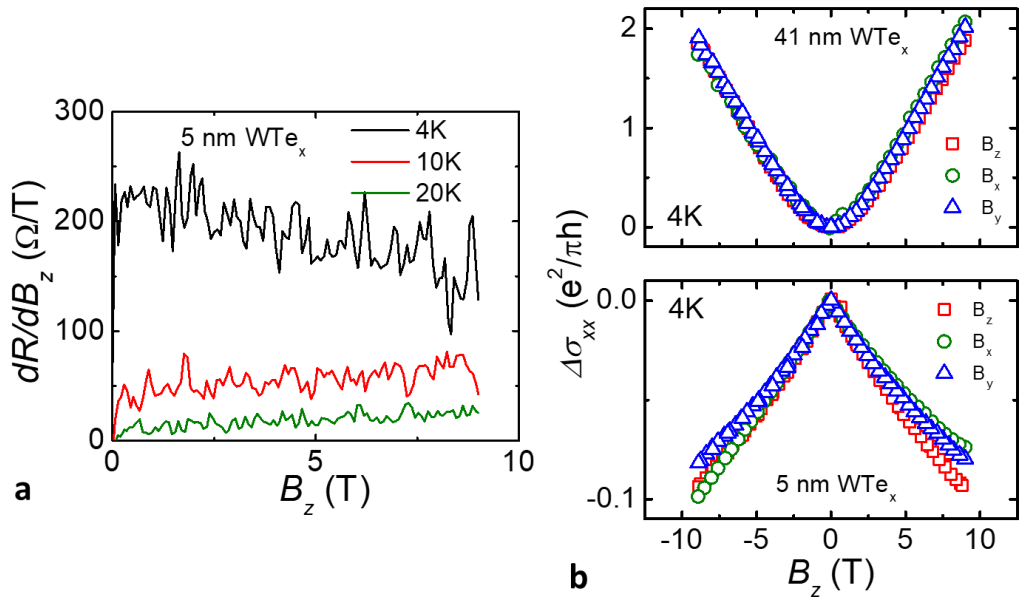

**Figure S4. Linearity analysis of WTe$_x$ magnetoresistance and WTe$_x$ magnetoresistance measured under different magnetic field orientations. a,** Derivative of magnetoresistance of the perpendicular magnetic field ($dR/dB_z$) as a function of $B_z$ for 5 nm WTe$_x$ sample. **b,** Change in sheet conductance ($\Delta\sigma_{xx}$) dependence on external magnetic fields for 5 nm thick and 58 nm thick WTe$_x$ films at 4 K, measured under perpendicular ($B_z$), longitudinal ($B_x$), and transverse ($B_y$) magnetic fields.

**Note S5. Analysis of magnetoresistance in 5 nm and 41 nm WTe$_x$ thin film**

Using the $B_\phi$ values in Figure 2**b**, we calculate the effective dephasing length $l_\phi = \sqrt{\hbar/4eB_\phi}$ (ranging from 18 nm at 4 K to 5.2 nm at 50 K) is larger than the WTe$_x$ thickness. In contrast, as shown in Figure S5**a--b**, fit of the MR curves based on a formula for the 3D system as proposed in [8] gives rise to $l_\phi$ and $p$ values that are not consistent with the 3D system model. The $B_\phi$ values data are fitted using $aT^{0.5}$-b function, with effective dephasing length $l_\phi$ ranging from 4.5 nm to 9.1 nm, which are mostly larger than the WTe$_x$ thickness of 5 nm. Moreover, as all known decoherence mechanisms in 3D give rise to $p > 1$ in the $B_\phi \sim T^p$ fit [9], the 5 nm WTe$_x$ film cannot be described using the 3D Weyl semimetal theory.

Using the $B_\phi$ values in Figure 2**b**, we calculate the effective dephasing length $l_\phi = \sqrt{\hbar/4eB_\phi}$ (ranging from 6 nm at 4 K to 3.5 nm at 50 K), which are much smaller than the WTe$_x$ thickness (41 nm). Meanwhile a fit of data in Figure 2**c** using the 2D equation [10] fails as shown in Figure S5**c**.

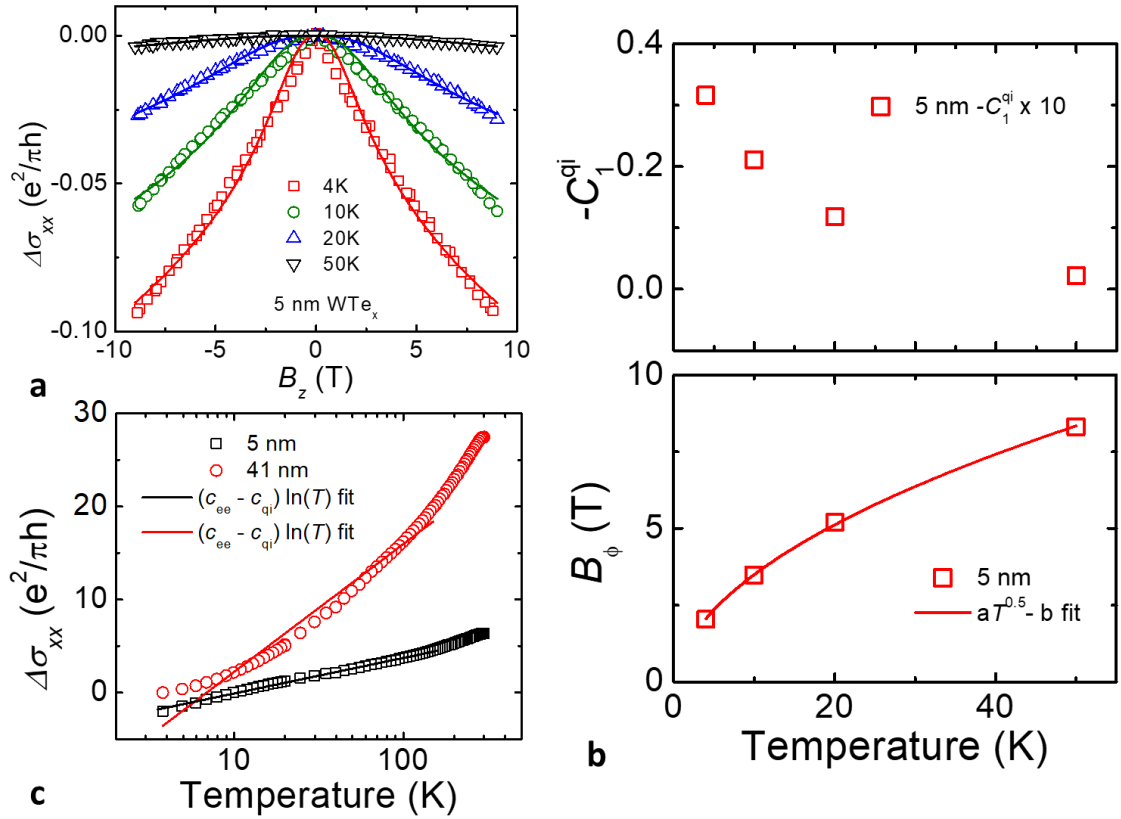

**Figure S5. a, b, Analysis of magnetoresistance in 5 nm WTe$_x$ thin film using an equation describing WL and WAL in 3D Weyl semimetals as proposed in** [8]**. a,** Change in sheet conductance ($\Delta\sigma_{xx}$) dependence on $B_z$, measured at 4, 10, 20, and 50 K. The fits using 3D WL/WAL formula deviate significantly from the raw data, especially at small field range for 4 and 10 K. **b,** Dependence of phase coherence characteristic field ($B_\phi$) and fit parameter ($-C_1^{qi}$) on temperature. **c, Analysis of 41 nm WTe$_x$ thin film using equation describing 2D massless Dirac fermions as proposed in** [10]**.** Change in sheet conductance ($\Delta\sigma_{xx}$) dependence on temperature. The 5 nm and 41 nm data below 150 K are fitted using $(c_{ee} - c_{qi})\ln(T)$ equation. The bad fit on the 41 nm data indicates that the 41 nm thick WTe$_x$ film cannot be described as a 2D massless Dirac fermion system.

**Note S6. Analysis of temperature dependent resistivity**

Four-point probe measurement on a fabricated Hall bar is employed to measure the resistivity of CoFeB(5)/MgO(2)/Ta(2), and WTe$_x$ ($t_{WTe_x}$)/CoFeB(4)/MgO(2)/Ta(2) heterostructures. We find that the 2 nm Ta is oxidized and does not conduct any current, as seen in Figure S9**a**. Hence, using a parallel resistor model, we can extract the WTe$_x$ resistivity dependence on temperature as shown in Figure S6**b**, using the CoFeB resistivity data at different temperatures, as shown in Figure S6**a**. Note that above 70K, the CoFeB resistivity shows a metallic behavior, while the low-temperature behavior can be attributed to weak localization [11]. Figure S6**b** shows a plot of resistivity both 5 nm and 41 nm samples in log scale as a function of $T^{-1/4}$. The linear dependence at T > 150 K ($T^{-1/4}$ < 0.285) follows the Mott's law for variable range hopping in 3D $\rho \propto \exp(T^{-1/4})$ [12]. While the low temperature resistivity data deviates from the high temperature data significantly. This indicates there is another conduction mechanism in these films.

Note that a fit solely based on thermal activation model using Arrhenius equation $\rho = \rho_0 \exp(E_a/k_B T)$, or a two-channel model consisting of a metallic channel and a thermal activation channel does not yield a good fit, especially in the low temperature range. Addition of the Arrhenius equation on top of the two-channel model used in the main text gives rise to an activation energy of several meV and $\rho_0$ that is at least 2 orders of magnitude smaller than $\rho_{bulk}$ from the VRH model, which thus can be neglected.

Figure S6**c** shows the conductance of the VRH and metallic channels for each sample while Figure S6**d** presents the evolution of fitting parameters as a function of $t_{WTe_x}$. The slight decrease in $\rho_{VRH}$ and $T_0$ suggests the increasing segregation disorder slightly enhances the bulk VRH conductivity (also see Figure S6**c**), while the drastic increase of $\rho_{metal}$ with $t_{WTe_x}$ indicates that increasing segregation disorder suppresses the metallic states significantly.

Here, the room temperature resistivity of the 5 nm WTe$_x$, i.e. 159 $\mu\Omega\ cm$, is smaller than previous reported 460-1400 $\mu\Omega\ cm$ for exfoliated WTe$_2$ flakes from 5L to 17L.[13] Considering that the semi-metallic channel consists of over 90% of conduction at room temperature as shown in Figure 3**b**, we speculate there are additional conduction channels arising in these amorphous WTe$_x$ films compared with crystalline counterparts.

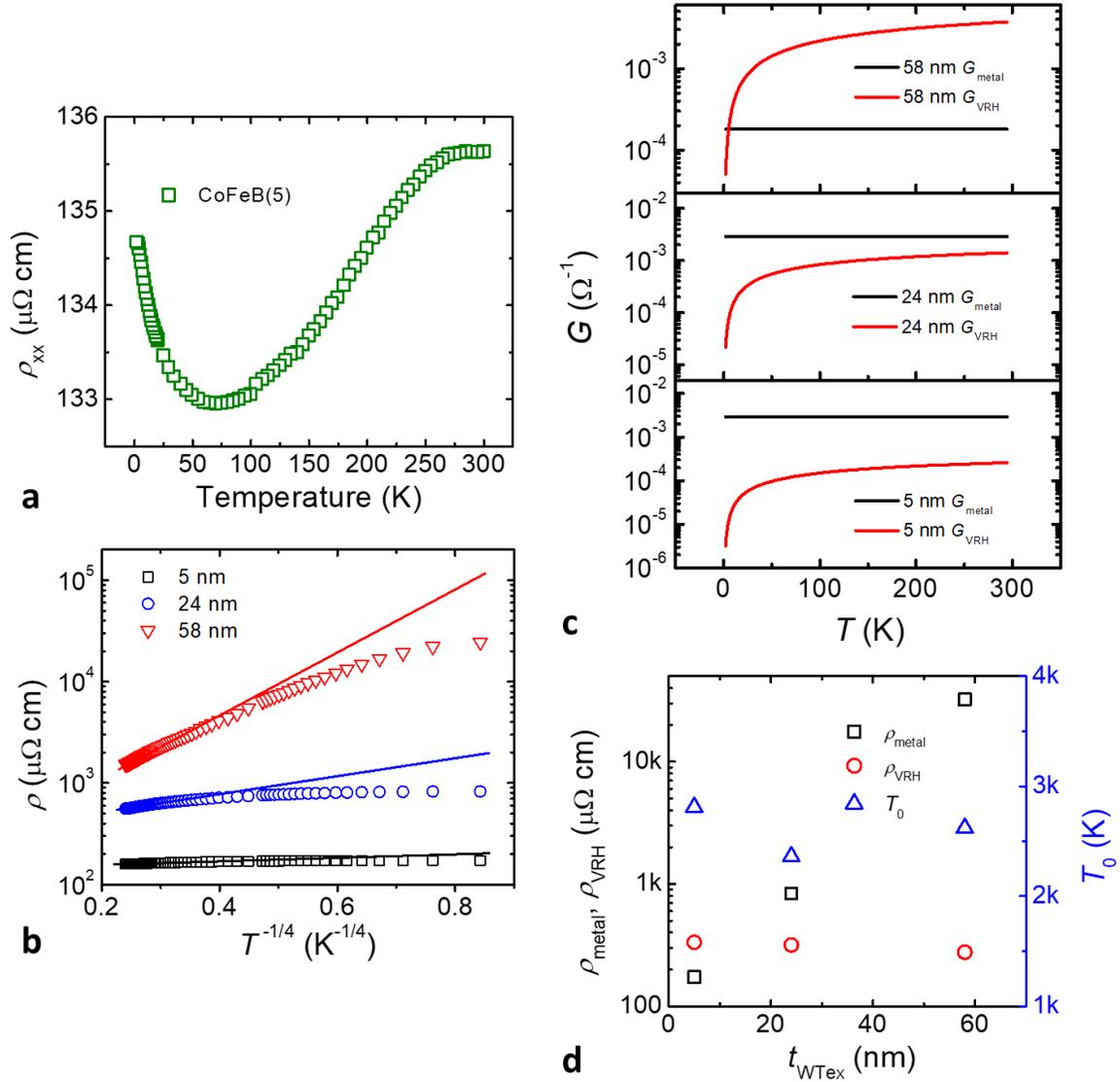

**Figure S6. Temperature dependent resistivity data analysis. a**, 5 nm CoFeB resistivity measured as a function of temperature. **b,** Resistivity ($\rho$) as a function of $T^{-1/4}$. **c,** Contribution of metallic states and VRH states to the total conductance for 5 nm, 24 nm, and 58 nm WTe$_x$ samples. **d,** Two-channel conduction model fit parameters $\rho_{metal}, \rho_{VRH}, T_0$ as a function of WTe$_x$ thickness.

Note that the room temperature resistivity of the 5-nm-thick WTe$_x$ in Figure 2 is 2590 $\mu\Omega\ cm$, which is much larger than 159 $\mu\Omega\ cm$ of 5-nm-thick WTe$_x$ in Figure 3 (or room temperature Figure S6**b** data). The main reason for this discrepancy is the 5-nm-thick WTe$_x$ in

Figure 2 is directly capped with MgO, while the 5-nm-thick WTe$_x$ in Figure 3 (or room temperature Figure S6**b** data) is capped with CoFeB/MgO. Hence the top surface of the WTe$_x$ 5 nm films is oxidized thus the resistivity is enhanced. Nevertheless, in the presence of surface oxidation, the observation of 2D Weyl fermion-like transport in the 5-nm-thick WTe$_x$ still holds. In addition, we did not observe any signatures of surface Fermi-arcs or Rashba 2D interface in WTe$_x$/CoFeB bilayers as discussed in the main text. Therefore, the oxidation of the surface of the 5-nm-thick WTe$_x$ does not alter its bulk-originated charge-to-spin conversion properties. On the other hand, the room temperature resistivity of the 41-nm-thick WTe$_x$ in Figure 2 is 1369 $\mu\Omega\ cm$, is similar to 1534 $\mu\Omega\ cm$ of 58-nm-thick WTe$_x$ in Figure 3 (or room temperature Figure S6**b** data). This is consistent with the above-mentioned 5-nm-thick WTe$_x$ result as the impact of a few nanometers of surface oxidation to the resistivity of a 41-nm-thick WTe$_x$ is negligible.

**Note S7. Analysis of ST-FMR data to extract charge-to-spin conversion efficiency**

As-deposited WTe$_x$/CoFeB(4)/MgO(2)/Ta(2) (all numbers in parenthesis hereafter are in the unit of nm) stacks were patterned into 40-$\mu$m-wide ($W$) and 60-$\mu$m-long ($L$) microstrips. By flowing RF current through the WTe$_x$/CoFeB bilayers, the oscillating current generates oscillating spin torques onto the CoFeB layer only including the damping-like and Oersted torques ($\tau_{DL}, \tau_{Oe}$). We have confirmed this and there is no field-like torque in Supplementary Note S13. When the RF current is in ferromagnetic resonance frequency, the CoFeB anisotropic magnetoresistance is also oscillating in magnitude, which when mixed with the RF current, generates a DC voltage. Here, the $H$ field is oriented with a $\phi$ angle of 45º with respect to the microstrip thus current flow direction. Using the following equation [14], we fit the mixing voltage using symmetric and asymmetric Lorentzian curves:

$$V_{mix} = -\frac{S\Delta^2}{\Delta^2+(H-H_0)^2} - \frac{A\Delta(H-H_0)}{\Delta^2+(H-H_0)^2} + C \quad \text{Equation 1}$$

where $S$ and $A$ are the coefficients for the symmetric and asymmetric Lorentzian functions centered at resonance field $H_0$ with a linewidth of $\Delta$, and $C$ is the offset. With the fitted curves, we can further calculate the charge-to-spin conversion efficiency $\xi_{ST}$ corresponding to the damping-like torque ($\tau_{DL}$) using the following equation:

$$\xi_{ST} = \frac{S}{A}\frac{e\mu_0 M_S t_{CoFeB} t_{WTe_x}}{\hbar}\sqrt{1+\frac{\mu_0 M_{eff}}{H}} \quad \text{Equation 2}$$

where $e$ is the electron charge, $\mu_0$ is the vacuum permeability, $\hbar$ is the reduced Planck's constant, $t_{CoFeB}$ ($t_{WTe_x}$) is the thickness of the CoFeB (WTe$_x$) layer, $M_S$ is the saturation magnetization of the CoFeB, $\mu_0 M_{eff}$ is the effective demagnetization field of the CoFeB layer, which includes both demagnetizing shape anisotropy and interfacial perpendicular magnetic anisotropy. Further, as shown in Figure S7**a-b**, we can obtain the $\mu_0 M_{eff}, \alpha$ values by fitting of resonance frequency as a function of the resonance field using the Kittel equation as below.

$$f_{res} = \frac{\gamma}{2\pi}\sqrt{H_{res}(H_{res}+4\pi M_{eff})} \quad \text{Equation 3}$$

$$\alpha = \gamma(\Delta - \Delta_0)/2\pi f_{res} \quad \text{Equation 4}$$

where $\gamma$ is the gyromagnetic ratio, $\Delta_0$ is the linewidth due to film inhomogeneity broadening,[15] $H_{res}$ ($f_{res}$) is the resonance field (frequency). Here as the CoFeB layer is rather thick (4.4 nm), we assume minimal contribution from interfacial perpendicular magnetic anisotropy in $\mu_0 M_{eff}$, thus $\mu_0 M_S = \mu_0 M_{eff} = 1.09\ T$.

We have also plotted the $S$ and $A$ values as a function of resonance frequency for the same device as used in Figure 4**a**. As shown in Figure S7**c**, no significant dependence is observed across 6 to 9 GHz. In addition, we calculate the skin depth of 5-58 nm WTe$_x$ with resistivity ranging from 160 – 1534 $\mu\Omega\ cm$ is all larger than 6 $\mu m$ for frequencies up to 10 GHz. Therefore, we believe there are minimal artefacts from skin-depth effects in our ST-FMR analysis.

Next, we analyze the contribution of spin pumping signal $V_{SP}$ from the CoFeB layer to the ST-FMR signal. This spin injection originated from the inverse spin Hall effect or inverse Rashba-Edelstein effect has the same field and angle dependence as the symmetric Lorentzian voltage $V_S^{ST-FMR}$ originated from the ST-FMR physics as mentioned above. Hence the total symmetric Lorentzian voltage $V_S = V_{SP} + V_S^{ST-FMR}$. To quantify the ratio of $V_{SP}/V_S^{ST-FMR}$, we employed a formula introduced in [16].

$$\frac{V_{SP}}{V_S^{ST-FMR}} = \frac{4Le^2 P \omega g_{eff}^{\uparrow\downarrow} M_S \lambda_{SD} \left(\frac{df}{dH}\right)_{H_{res}} t_{WTe_x}^2 t_{CoFeB} \sigma_{WTe_x} \tanh\frac{t_{WTe_x}}{2\lambda_{SD}}}{W \frac{dR}{d\phi} \gamma \hbar \Delta (t_{WTe_x}\sigma_{WTe_x} + t_{CoFeB}\sigma_{CoFeB})^2 [1-\text{sech}\frac{t_{WTe_x}}{\lambda_{SD}}]} \qquad \text{Equation 5}$$

where $P = \dfrac{2\omega\left[\gamma\mu_0 M_{eff} + \sqrt{(\gamma\mu_0 M_{eff})^2 + (2\omega)^2}\right]}{(\gamma\mu_0 M_{eff})^2 + (2\omega)^2}$ is the ellipticity correction factor to the DC voltage due to the ellipticity of the magnetization precession, $\omega$ is the ST-FMR frequency, $\gamma$ is the gyromagnetic ratio, $g_{eff}^{\uparrow\downarrow}$ is the effective interfacial mixing conductance, $\lambda_{SD}$ is the spin diffusion length in WTe$_x$, $\frac{dR}{d\phi}$ is the derivative angular dependence of the WTe$_x$/CoFeB strip resistance at $\phi = 45°$, $\sigma$ refers to the electrical conductivity.

We take the $\mu_0 M_S = \mu_0 M_{eff}$, $\Delta$, $\left(\frac{df}{dH}\right)_{H_{res}}$ values at $\omega = 6$ GHz as a representative frequency for the following calculations. For $g_{eff}^{\uparrow\downarrow}$ we take the $3.48 \times 10^{19} m^{-2}$ value which is consistent with experimental values from several HM/CoFeB bilayers.[17] The conductivity values of $WTe_x$ and CoFeB layers are taken from Supplementary Note S6. Here, the large variation of $\rho_{WTe_x}$ with different $t_{WTe_x}$, which can be attributed to the increase of segregation disorder as discussed in the main text, makes the simple drift-diffusion model not valid anymore.

To estimate the spin diffusion length $\lambda_{SD}$ in $WTe_x$, we refer to the relationship between $\xi_{ST}$ and $\lambda_{SD}$ in exfoliated $WTe_2$ flakes as reported in Figure S10 of reference [18]. We can find that for a $\xi_{ST}$ of 0.1 – 0.5 as found in $WTe_x$, the $\lambda_{SD}$ is in the range of 0.2 – 0.7 nm. This is also consistent with $\lambda_{SD}$ of 0.3 – 1 nm in amorphous heavy metals, [19] as well as $\lambda_{SD}$ around 0.6 nm for Pt in the dirty metal regime. [20] Using $\lambda_{SD}$ of 0.2 and 0.7 nm, we calculate $V_{SP}/V_S^{ST-FMR}$ to be around less than 0.06 as shown in Figure S7**e**. Hence, we can safely ignore the contribution from spin pumping in our ST-FMR results.

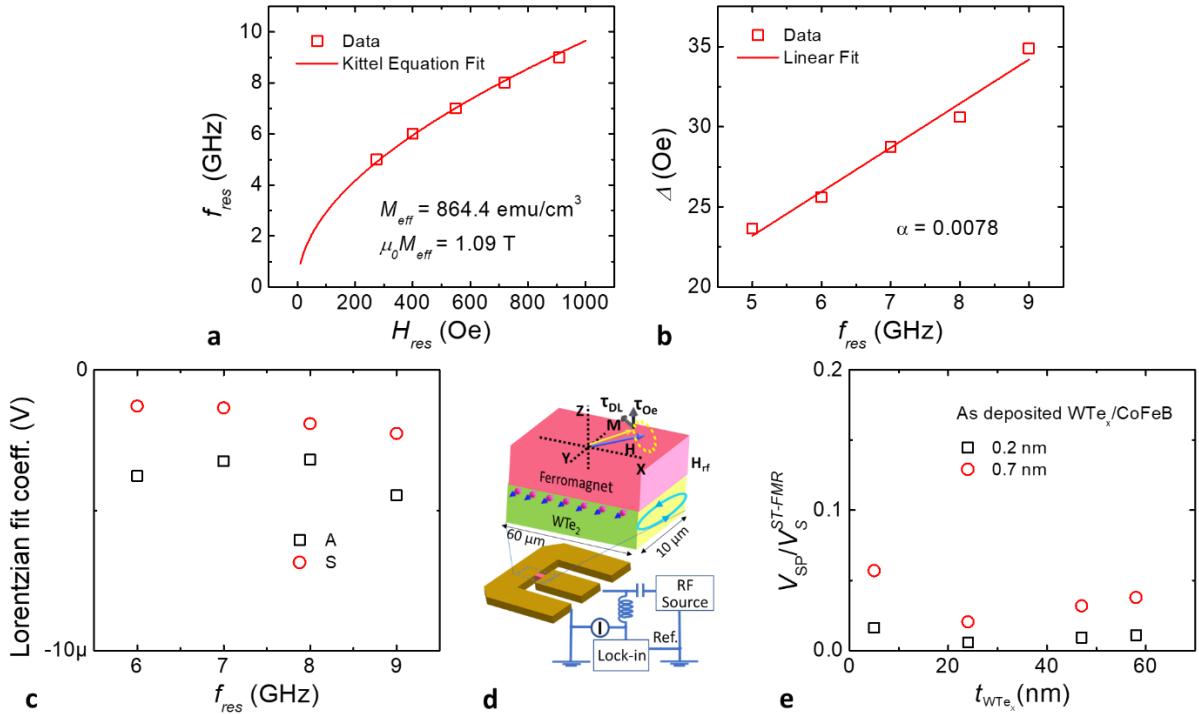

**Figure S7. ST-FMR resonance related data analysis. a,** Resonance frequency ($f_{res}$) as a function of the resonance field ($H_{res}$) fitted by Kittel equation to extract the effective demagnetization ($M_{eff}$). **b,** Resonance peak linewidth (Δ) as a function of resonance frequency ($f_{res}$) fitted by a linear fit to extract the damping constant $\alpha$. **c,** Symmetric and asymmetric Lorentzian coefficients (S/A) as a function of resonance frequency for the 58-nm-thick $WTe_x$

device in Figure 4a. **d,** Schematic of spin-torque ferromagnetic resonance (ST-FMR) experimental setup using fabricated WTe$_x$($t_{WTe_x}$)/CoFeB(4)/MgO(2)/Ta(2) microstrips. The damping-like and Oersted-field torques are labeled as $\tau_{DL}, \tau_{Oe}$ respectively. **e,** Calculated ratio of spin pumping voltage $V_{SP}$ over the symmetric Lorentzian voltage $V_S^{ST-FMR}$ from ST-FMR origin. Data for $\lambda_{SD}$ of 0.2 and 0.7 nm are plotted.

**Note S8. Raman spectrum and anomalous Hall data of WTe$_x$/CoFeB and WTe$_x$/Mo/CoFeB heterostructures**

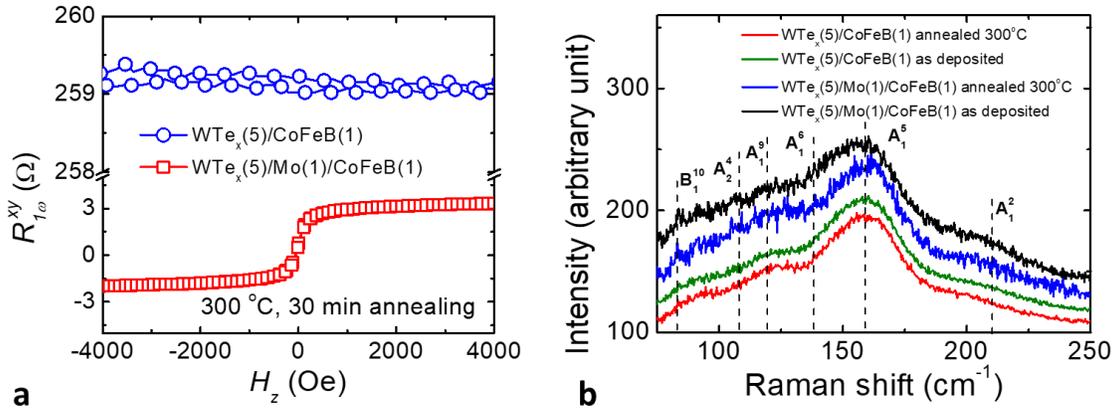

**Figure S8. Raman spectrum and anomalous Hall data of WTe$_x$/CoFeB and WTe$_x$/Mo/CoFeB heterostructures. a,** Raman spectrum of WTe$_x$(5)/CoFeB(1) and WTe$_x$(5)/Mo(1)/CoFeB(1) heterostructures before and after annealing at 300°C for 30 minutes. **b,** Anomalous Hall data as a function of the perpendicular magnetic field ($H_z$) of WTe$_x$(5)/CoFeB(1) and WTe$_x$(5)/Mo(1)/CoFeB(1) heterostructures after annealing at 300°C for 30 minutes. All films were sputtered on SiO$_2$/MgO(2) and capped with MgO(2)/Ta(2) films, the numbers in parenthesis have the unit of nm.

**Note S9. High resolution TEM and EDS element distribution of MgO(2)/WTe$_x$(5)/Mo(1)/CoFeB(1)/MgO(2) heterostructure**

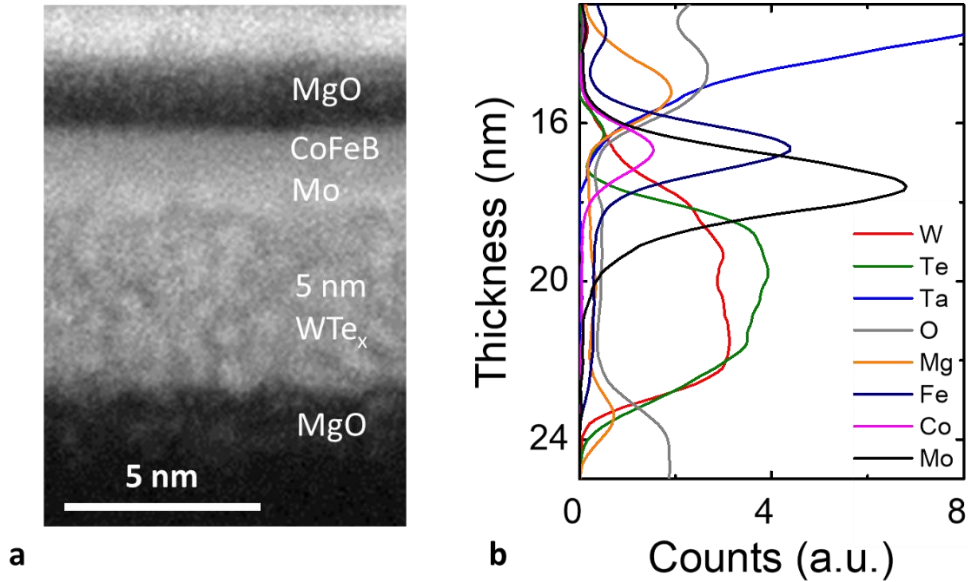

**Figure S9. High resolution TEM and EDS element distribution of MgO(2)/WTe$_x$(5)/Mo(1)/CoFeB(1)/MgO(2) heterostructure. a,** Cross section scanning transmission electron microscopy (STEM) image. **b,** Energy dispersive X-ray spectroscopy (EDS) element distribution profile. Note that from Figure 4**b** and Figure 5**c** the damping-like SOT from the WTe$_x$ layer has a positive sign which is consistent with exfoliated WTe$_2$ [21] and opposite from that of W. Henceforth, the large positive SOT cannot be ascribed to the W-rich regions which is adjacent to the CoFeB layer and overlapping with the Mo region as shown in Figure S9**b**. The SOT efficiency from the WTe$_x$ layer might be even larger after considering a negative-sign SOT contribution from the diffused W-rich regions.

**Note S10. Analysis of second-harmonic Hall data to extract charge-to-spin conversion efficiency**

Like the ST-FMR technique, the varying Hall resistance driven by the oscillating SOT multiplied by the varying AC current gives rise to a second harmonic Hall resistance $R_{2\omega}^{xy}$. In general, for an in-plane magnetized CoFeB layer, which has very small in-plane anisotropy, its magnetization will saturate along the external in-plane magnetic field direction. The first and second harmonic resistance ($R_{1\omega}^{xy}, R_{2\omega}^{xy}$) can then be expressed using the following equations [22-24]

$$R_{1\omega}^{xy} = R_P \sin^2\theta \sin 2\phi + R_A \cos\theta \qquad \text{Equation 6}$$

$$R_{2\omega}^{xy} = R_P \frac{H_{FL}+H_{Oe}}{|H|}\cos 2\phi \cos\phi + \frac{1}{2}\left(R_A \frac{H_{DL}}{|H|-H_K} + R_{TE}\right)\cos\phi \qquad \text{Equation 7}$$

where $R_P, R_A, R_{TE}$ refer to the planar Hall, anomalous Hall, and thermoelectric (including anomalous Nernst and spin Seebeck effects) resistance respectively, $H_{FL}, H_{Oe}, H_{DL}$ refer to field-like, Oersted, and damping-like field respectively, $H_K$ refers to the effective demagnetization field, $\theta$ is the angle of magnetization with respect to the $z$-axis, and $\phi$ is the angle of magnetization with respect to the $x$-axis or the current flow direction.

First, we measured $R_{2\omega}^{xy}$ as a function of in-plane magnetic field angle $\phi$ as shown in Figure S10**a**. For the $H$ field magnitude ranging from 130 Oe to 6 Oe, we can fit the $R_{2\omega}^{xy} - \phi$ data using a cosine function very well, indicating there is a minimal contribution of field-like and Oersted field in our heterostructure. Then, we fix the in-plane $\phi$ angle to 45° and vary the AC current amplitude. As shown in Figure S10**a**, these data gathered can then be fitted using Equation 6 excluding the $\cos 2\phi \cos\phi$ term. Further, in the $\cos\phi$ term, only the damping-like field depends on the external magnetic field while the thermoelectric effects do not. By plotting $R_{2\omega}^{xy}$ as a function of $1/(|H|-H_K)$ as shown in Figure S10**b**, we can determine that the thermal contributions $R_{TE}$ are also negligible. Note that the $R_A$ and $M_S$ values can be obtained by measuring the Hall resistance and magnetization as a function of the perpendicular magnetic field, as shown in Figure S10**c-d**. While the effective demagnetization field $H_K$ is obtained to be around - 478 Oe from Figure S10**c**.

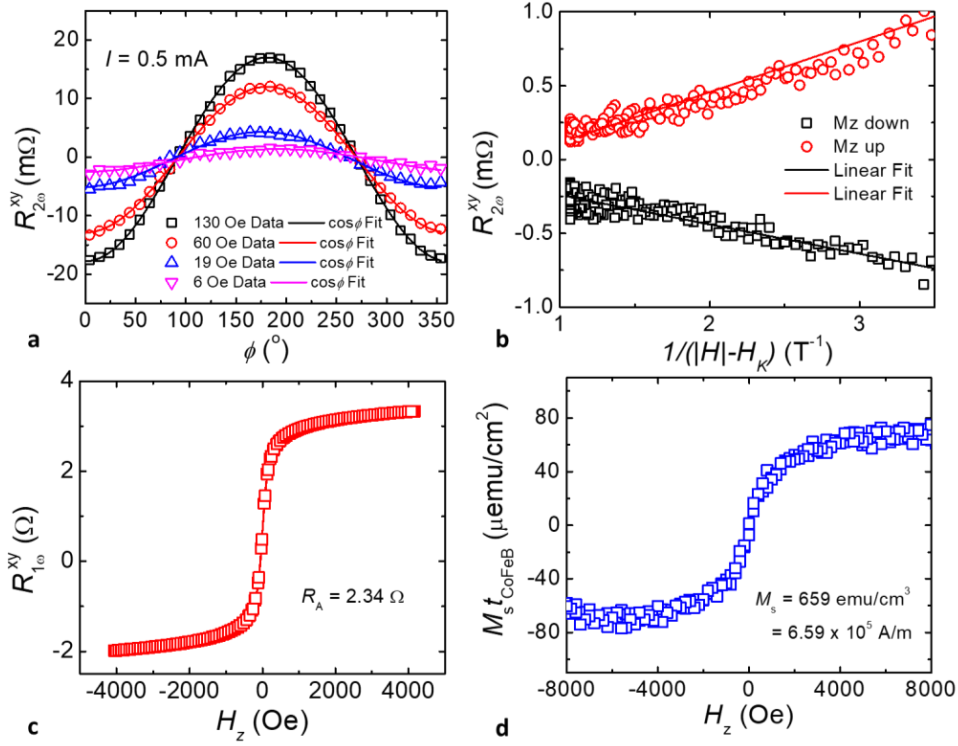

**Figure S10. Second harmonic measurements of spin-orbit-torques in WTe$_x$(5)/Mo(1)/CoFeB(1)-based heterostructure. a**, In-plane magnetic field angle ($\phi$) dependence of second harmonic Hall resistance ($R_{2\omega}^{xy}$) as a function of magnetic field $H$ strength with $I = 0.5\ mA$, which are fitted using a $\cos\phi$ function. **b,** $R_{2\omega}^{xy}$ dependence on the inverse of the in-plane effective magnetic field (1/(|H|-H$_K$)) graph with $\phi = 45°$ and $I = 0.5\ mA$, which are linearly fitted to extract the thermal contribution to the $R_{2\omega}^{xy}$ value. **c,** Anomalous Hall resistance ($R_{1\omega}^{xy}$) dependence on out-of-plane magnetic field ($H_z$). **d,** Product of saturation magnetization ($M_S$) and CoFeB thickness ($t_{CoFeB}$) dependence on $H_z$.

**Note S11. Analysis of WTe$_x$ resistivity in WTe$_x$/Mo/CoFeB heterostructures**

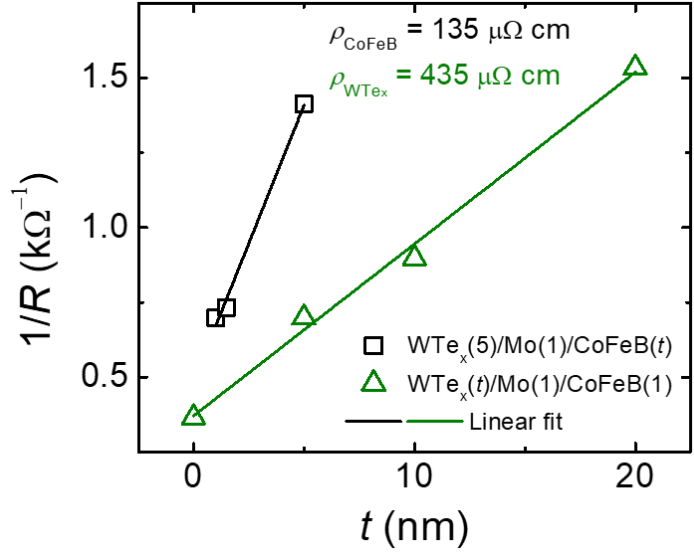

**Figure S11. Extraction of CoFeB, Mo, and WTe$_x$ thin film resistance.**
MgO(2)/WTe$_x$(5)/Mo(1)/CoFeB($t$)/MgO(2)/Ta(2), and
MgO(2)/WTe$_x$($t$)/Mo(1)/CoFeB(1)/MgO(2)/Ta(2) heterostructures resistance are measured using four-point measurement on a $10 \times 40~\mu m$ Hall bar.

The material (CoFeB/WTe$_x$) resistivity is extracted from the slope of the Hall bar conductivity ($1/R$) as a function of the variable material (CoFeB/WTe$_x$) thickness $t$. In this linear fit, the intercept corresponds to the value of the two layers left when the variable material (CoFeB/WTe$_x$) thickness vanishes to zero. Then we use the fitted CoFeB and WTe$_x$ resistivity values and the two intercept values to derive the Mo resistivity values to be 126/135 $\mu\Omega~cm$. Here, the CoFeB resistivity is consistent with data in Figure S7**a**. Using the CoFeB resistivity, we also confirm the Mo resistivity is 139 $\mu\Omega~cm$ and 136 $\mu\Omega~cm$ in MgO(2)/Mo(1)/CoFeB(1)/MgO(2)/Ta(2) and MgO(2)/Mo(6)/CoFeB(1)/MgO(2)/Ta(2) heterostructures, respectively. This is consistent with the 126/135 $\mu\Omega~cm$ values obtained above, thus confirming the validity of this parallel resistor model to extract each layer's resistivity. This also confirms that the Mo phase does not change in the thickness range from 1 nm to 6 nm.

Compared with literature, the CoFeB resistivity and WTe$_x$ resistivity is consistent with previous reported 170 $\mu\Omega~cm$ for sputtered CoFeB [25] and 460 $\mu\Omega~cm$ for exfoliated WTe$_2$ flakes [13]. While the Mo resistivity is larger than previous reported values of 85 $\mu\Omega~cm$.[26] Note that the resistivity of WTe$_x$ 435 $\mu\Omega\ cm$ is also larger than the 159 $\mu\Omega\ cm$ value obtained from 5 nm WTe$_x$/CoFeB bilayers as in Figure 4**b**. We think the increased WTe$_x$ and Mo resistivities in the WTe$_x$/Mo/CoFeB samples may be attributed to the strong intermixing of Mo and WTe$_x$ after thermal annealing, thus strong interface scattering, as shown in Figure S9**b**.

**Note S12. Use of second harmonic measurements and ST-FMR measurements in WTe$_x$/CoFeB and WTe$_x$/Mo/CoFeB heterostructures**

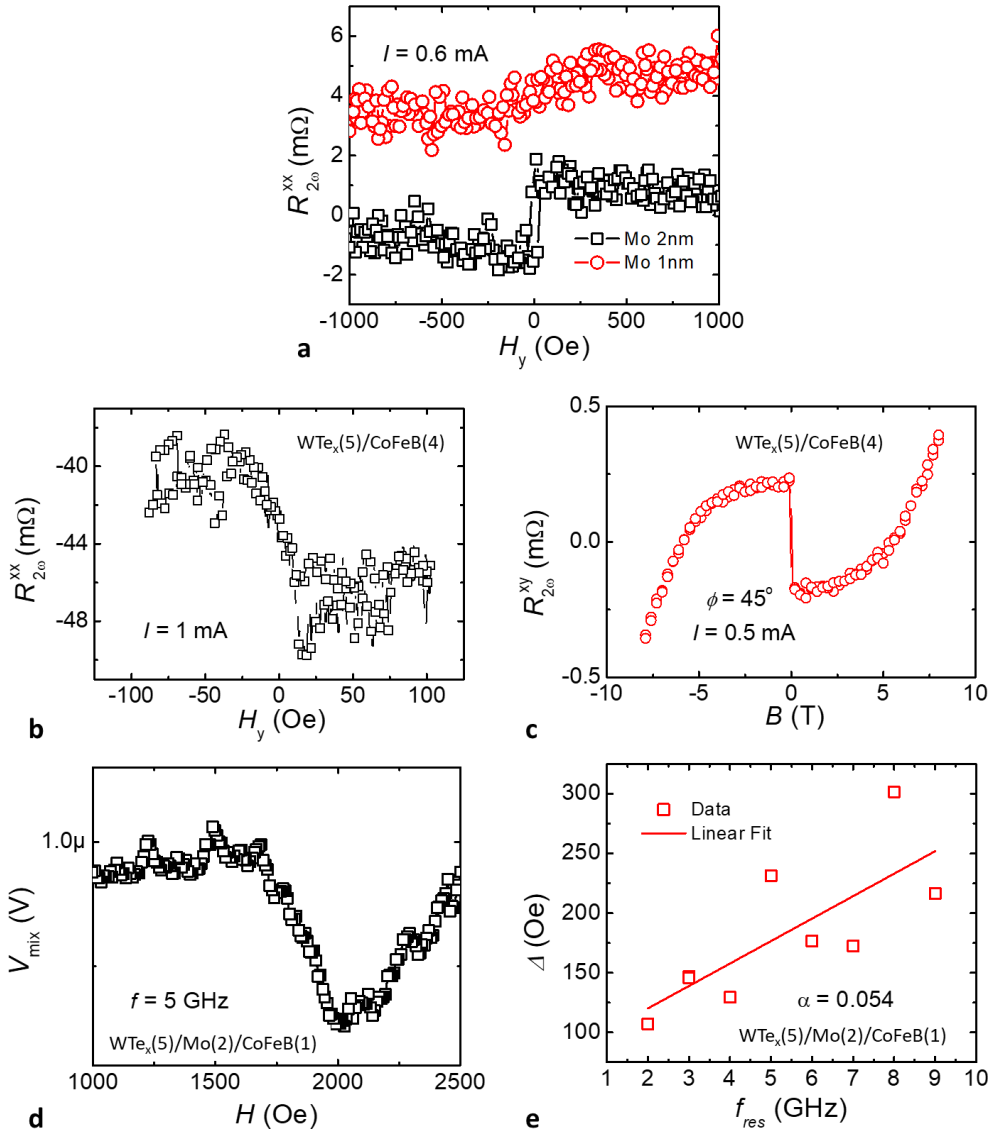

**Figure S12. Second harmonic and ST-FMR results in WTe$_x$/CoFeB and WTe$_x$/Mo/CoFeB heterostructures. a,** Second harmonic longitudinal Hall resistance in WTe$_x$/Mo/CoFeB heterostructures. Data were gathered from MgO(2)/WTe$_x$(5)/Mo(1 or 2)/CoFeB(1)/MgO(2)/Ta(2) heterostructure when the magnetic field ($H_y$) is swept in the transverse direction ($y$). The AC current amplitude in the Hall bar channel is 0.6 mA. **b,** Second harmonic longitudinal Hall resistance in WTe$_x$(5)/CoFeB(4) heterostructure. **c,** Second harmonic transverse Hall resistance as a function of in-plane magnetic field in WTe$_x$(5)/CoFeB(4) heterostructure. **d,** Mixing voltage

obtained in ST-FMR measurement as a function of applied magnetic field $H$ strength for WTe$_x$(5)/Mo(2)/CoFeB(1) heterostructure measured at 5 GHz RF excitation. **e,** Extraction of damping constant from ST-FMR linewidths obtained in WTe$_x$(5)/Mo(2)/CoFeB(1) heterostructure.

Here, we explain the reasons of choosing WTe$_x$/Mo/CoFeB samples for second harmonic, and WTe$_x$/CoFeB samples for ST-FMR measurements.

First, the 2$^{nd}$ transverse or longitudinal Harmonic measurements results on the WTe$_x$(5 nm)/CoFeB(4 nm) samples cannot be used for accurate SOT analysis. As shown in Figure S12**b**, the longitudinal resistance as a function of in-plane field $H_y$ does not exhibit a positive coercivity, hence a current-driven SOT switching experiment is not possible. Similar results are also shown for other WTe$_x$/CoFeB bilayers in Figure 3 and 4. Then as shown in Figure S12**c**, the transverse resistance as a function of in-plane field $H$ ($\phi = 45°$) shows a very large signal that is proportional to the external magnetic field up to 9T. This large linear background prevents the accurate analysis of the damping like torque in this sample. While in Figure 5, the magnetic field magnitude is at most around 0.7T, the impact of the large linear background can be neglected. Similar results are also shown for other WTe$_x$/CoFeB bilayers in Figure 3 and 4. Lastly, from previous studies of SOT in exfoliated WTe$_2$ flakes, this large signal has been attributed to the chiral-anomaly-induced giant second harmonic plan Hall resistance.[21, 27] Second, the ST-FMR results on the WTe$_x$(5nm)/Mo(2)/CoFeB(1nm) sample cannot be used for accurate SOT analysis either. As shown in Figures S12**d-e** below, we have measured the damping of 1 nm CoFeB to be around 0.054, which is around 10 times larger than that from WTe$_x$/CoFeB bilayers. This leads to a set of data that is too noisy for accurate ST-FMR analysis. The very large damping in the 1 nm CoFeB could be attributed to its small thickness and interdiffusion of Mo and CoFeB after thermal annealing.

**Note S13. Analysis of thermoelectric contribution to second harmonic longitudinal resistance data and effect from field-like and Oersted field during SOT switching**

We first find contribution from Oersted field torque in the in-plane angle-dependent second harmonic Hall resistance data of the WTe$_x$(5)/Mo(2)/CoFeB(1)/MgO(2)/Ta(2) device that we carry out switching experiment in Figure 5**b**. First, we fit the in-plane $\phi$ angular dependence of the $R_{2\omega}^{xy}$ signal as shown in Figure S13**a**. Note that for $H$ larger than 78 Oe, a simple $\cos\phi$ fit is sufficient, whereas for $H$ of 5.5 Oe, the addition of the $\cos 2\phi \cos\phi$ term is required. This indicates that the effect of field-like torque and Oersted field is very small, thus only appearing when external field $H$ is small. As shown in Figure S13**b**, the fitted $H_{FL+Oe}$ is plotted as a function of the AC current amplitude $I$. A linear fit yields a slope of 0.154 $Oe/mA$, corresponding to effective efficiency of 0.0057 using the same calculation method of $\xi_{ST}$. Note that compared with the 1-nm-thick Mo case as shown in Figure S10**a** where no Oersted-field contribution is seen when external field is 6 Oe, the use of 2-nm-thick Mo here which is more conductive indeed enhances the Oersted-field torque as shown in Figure S13**a**.

As the coercivity of the CoFeB layer for SOT switching in Figure 5**a** is rather small (1.5 Oe and -1.2 Oe where $R_{2\omega}^{xy}$ passes around -5 $m\Omega$), it is important to consider the effects from field-like and Oersted field. As with the critical switching current in Figure 5**b** (3.8 mA and -2.8 mA), according to the slope from the linear fit, $H_{FL+Oe}$ is calculated to be 0.59 Oe and -0.43 Oe respectively. These values are roughly 1/3 of the correspondingly coercivity of the CoFeB layer. In addition, we note that the coercivity found in Figure 5**a** is smaller than the actual coercivity because an AC current of 0.7 mA is applied during the measurement. Hence, the actual coercivity should be larger than that stated above. Meanwhile, the sign of the $H_{FL+Oe}$ is the same of $H_{DL}$. Hence, we conclude that the field-like and Oersted field assists the SOT-driven switching process.[28]

Next, we analyze the relative contribution of Oersted field and field-like SOT field. We use the simple Ampere's law to calculate $H_{Oe} = \mu_0 I_{WTe_x+Mo}/2w$, where $I_{WTe_x+Mo}$ is the current running through the WTe$_x$/Mo bilayers based on the parallel resistor model in Figure S11, and $w$ is the width of the Hall bar device. The $H_{Oe}$ values are determined to be 1.9 Oe and -1.4 Oe for the critical switching current in Figure 5**b**. However, this simple calculation cannot capture the gradience of Oersted field on the CoFeB layer or the reduction of actual Oersted field due to the permeability of the CoFeB layer. Based on a simulation on WTe$_2$/NiFe bilayers[27], the actual Oersted field can be 25-66% of the calculated value from the simple Ampere's law equation. Hence, we believe that the actual $H_{Oe}$ exerted on the CoFeB layer is consistent with the $H_{FL+Oe}$ values measured experimentally. This dominating Oersted field and negligible field-like SOT field is also consistent with previous reports on WTe$_2$ flakes.[21, 27, 29]

Then using the same method as described in Supplementary Note 11, i.e., by plotting $R_{2\omega}^{xy}$ as a function of $1/(|H|\text{-}H_K)$, we further fit the thermal electric contribution to the second harmonic Hall resistance ($R_{2\omega}^{xy-TE}$) for different AC current amplitude as shown in Figure S13**c**. Note that we ignore the contribution from Oersted field in this analysis because the region of $R_{2\omega}^{xy}$ values where the effect of Oersted field emerges (when $|H| < 78$ Oe) is not used to extract the $R_{2\omega}^{xy-TE}$ values ($R_{2\omega}^{xy}$ values with $|H| > 800$ Oe are used in extracting $R_{2\omega}^{xy-TE}$ values). By scaling the $R_{2\omega}^{xy-TE}$ with the longitudinal dimension ($L = 40$ μm) over transverse dimension ($W = 10$ μm) of the Hall bar, [30, 31] we obtain the thermal contribution to the total $R_{2\omega}^{xx}$ ($R_{2\omega}^{xx-TE}$) for different AC current amplitude as shown in Figure S13**d**. Note that we can also fit the $R_{2\omega}^{xy}$ dependence on external magnetic field $H$ as shown in Figure S13**c** with purely damping-like SOT contribution using Equation 6.

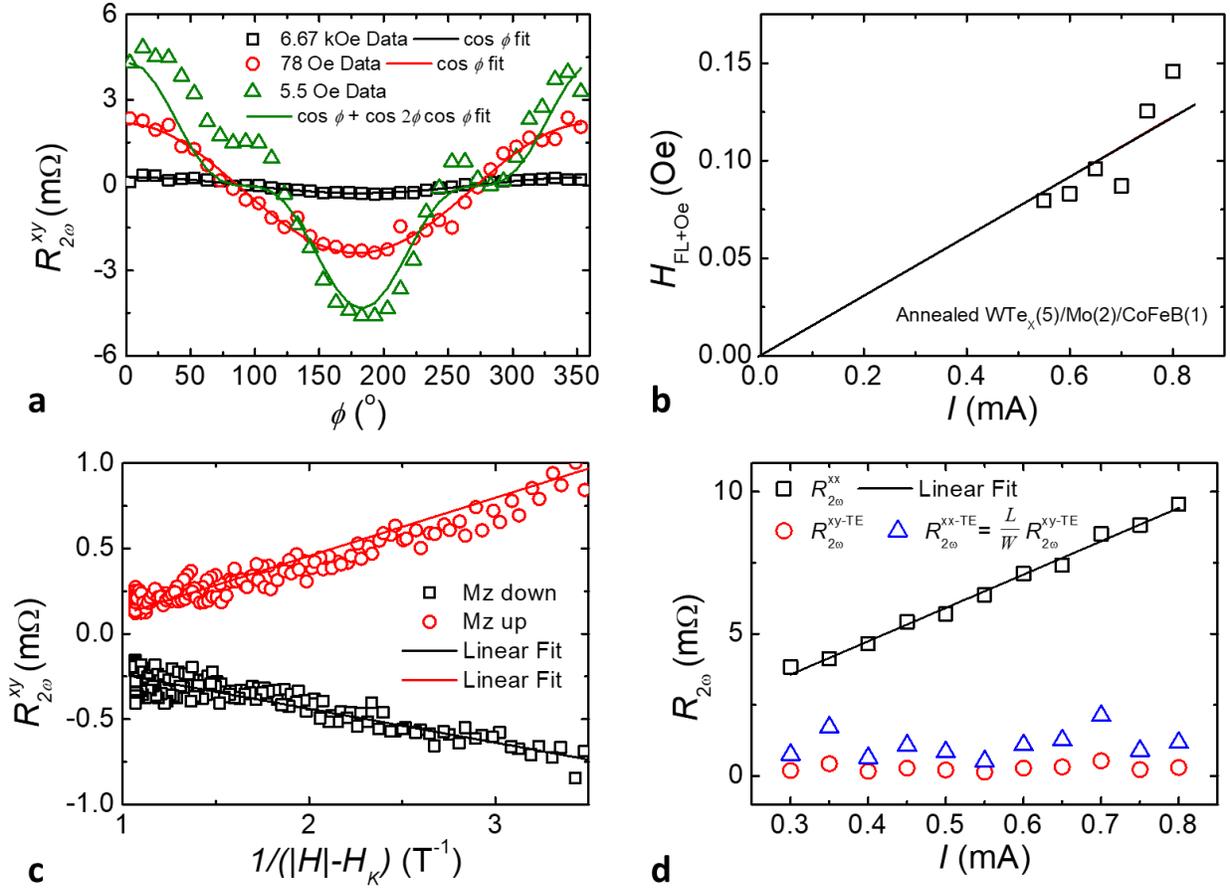

**Figure S13. Second harmonic measurements of spin-orbit-torques in WTe$_x$(5)/Mo(2)/CoFeB(1) heterostructure. a**, In-plane magnetic field angle ($\phi$) dependence of second harmonic Hall resistance ($R_{2\omega}^{xy}$) as a function of different magnetic field $H$ strengths with $I = 0.5\ mA$, which are fitted using a $\cos\phi$ and $\cos 2\phi \cos\phi$ function. **b,** Effective field from field-like and Oersted field $H_{FL+Oe}$ as a function of AC current amplitude $I$. The fit is a linear fit with intercept at zero. **c,** $R_{2\omega}^{xy}$ dependence on the inverse of the in-plane effective magnetic field $(1/(|H|-H_K))$ graph with $\phi = 45°$ and $I = 0.5\ mA$, which are linearly fitted to extract the thermal contribution to the $R_{2\omega}^{xy}$ value. **d,** Total $R_{2\omega}^{xx}$, thermal contribution to $R_{2\omega}^{xx}$ ($R_{2\omega}^{xx-TE}$), and thermal contribution to $R_{2\omega}^{xy}$ ($R_{2\omega}^{xy-TE}$) as a function of AC current amplitude.

**Note S14. Second harmonic USMR measurements of Pt(4.7)/CoFeB(2) heterostructure**

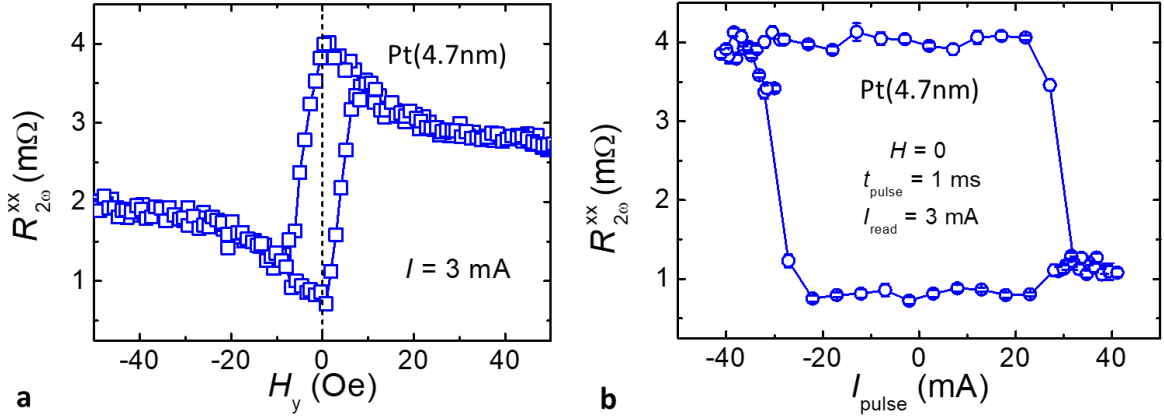

**Figure S14. Second harmonic USMR measurements of Pt(4.7)/CoFeB(2) heterostructure. a,** $R_{2\omega}^{xx}$ as a function of the in-plane magnetic field along $y$-axis ($H_y$) under an AC current amplitude of $I = 3\ mA$. **b,** $R_{2\omega}^{xx}$ as a function of pulse current amplitude $I_{pulse}$ under zero external field. The current pulse width is 1 ms. The read AC current amplitude is 3 mA. Data are measured in as-deposited Pt(4.7)/CoFeB(2)/MgO(1.5)/Ta(2) heterostructure without any annealing.

**Note S15. Analysis of out-of-plane damping-like torque**

First, from Figure S10**a**, and S13**a** the good fits of second harmonic resistance $R_{2\omega}^{xy}$ as a function of in-plane angle $\phi$ using a combination of $\cos\phi$ and $\cos 2\phi \cos\phi$ fits indicate that there is no out-of-plane damping-like torque, which follows a $\cos 2\phi$ dependence.[39]

Second, we plot ST-FMR data for both the positive and negative field as shown in Figure S15. If there were an out-of-plane damping-like torque, the asymmetric component will differ in magnitude for the different magnetic field polarities.[21, 39] This contrasts with what we observe in Figure S15 where the asymmetric component magnitudes are almost same regardless of the magnetic field polarity.

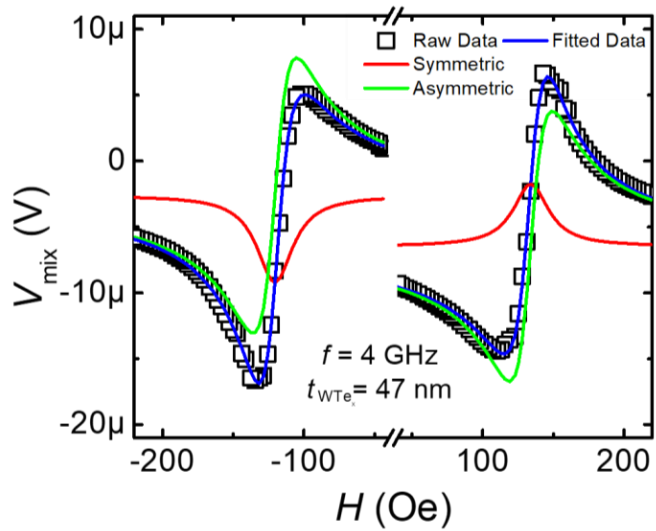

**Figure S15. ST-FMR measurements of WTe$_x$/CoFeB heterostructure. a**, Representative mixing voltage as a function of applied magnetic field $H$ strength for WTe$_x$ thickness of 47 nm measured at 4 GHz RF excitation. The raw data from both positive and negative magnetic fields are being fitted by a sum of symmetric and asymmetric Lorentzian.

**Table S1. Comparison of USMR strength $\Delta R_{2\omega}^{xx}/JR_{xx}$ with published works.** SF refers to spin-flip USMR originating from spin-dependent electron scattering between SOT spin current and magnons in the ferromagnet, SD refers to spin-dependent USMR arising from spin-dependent electron scattering between SOT spin currents and the magnetization at the interface or bulk, [32] SHE refers to spin Hall effect, ISGE refers to inverse spin galvanic effect, *Ultrahigh USMR strength appears as non-uniform and hysteretic data at low magnetic fields. **Surface spin-polarized carriers interaction with magnetic dopants.

| Material Stack | $\Delta R_{2\omega}^{xx}$ (m$\Omega$) | $R_{xx}$ ($\Omega$) | $\Delta R_{2\omega}^{xx}/R_{xx}$ (ppm) | $J$ (MA/cm$^2$) | $\Delta R_{2\omega}^{xx}/JR_{xx}$ (ppm MA$^{-1}$cm$^2$) | Origin | Reference |
|---|---|---|---|---|---|---|---|
| WTe$_x$/Mo/CoFeB | 17 | 954 | 17.8 | 0.42 | 42.4 | SF | This work |
| Bi$_2$Se$_3$/CoFeB (150K) | 1.0 | 733 | 1.4 | 0.67 | 6.1 | SD | 30 |
| Pt/Co | 15.0 | 177 | 84.8 | 10 | 8.5 | SF+SD | 31 |
| Ta/Co | 11.0 | 574 | 19.2 | 10 | 1.9 | SD | 31 |
| W/Co | 18.6 | 572 | 32.5 | 10 | 3.3 | N/A | 33 |
| Ta/Co/Pt/Ta | 102 | 284 | 360 | 50 | 7.2 | SF+SD | 32 |
| Ta/Co/Pt/Ta* | 350 | 284 | 1232.4 | 50 | 24.6 | SF+SD | 32 |
| Ta/CoFe/Pt/CoFe | | | 8.12 | 1.16 | 7.0 | SD | 34 |
| YIG/Bi$_2$Se$_3$ (300K) | | | 2.65 | 1.06 | 2.5 | SD | 35 |
| YIG/Bi$_2$Se$_3$ (150K) | 26.9 | 2595 | 10.38 | 0.44 | 23.6 | SD | 35 |
| Ga0.91Mn0.09As/ Ga0.97Mn0.03As (150K) | 2000 | 1720 | 1163 | 0.75 | 1550 | SHE+ISGE | 36 |
| Cr$_x$(Bi$_{1-y}$Sb$_y$)$_{2-x}$Te$_3$/ (Bi$_{1-y}$Sb$_y$)$_2$Te$_3$ (2K) | 57000 | 14000 | 4071 | 0.005 | $8.14 \times 10^5$ | SF | 37 |
| Cr$_{0.16}$(Bi$_{0.54}$Sb$_{0.38}$)$_2$Te$_3$/(Bi$_{0.5}$Sb$_{0.5}$)$_2$Te$_3$ (1.9K) | 36000 | 13300 | 2707 | 0.0003 | $9.02 \times 10^6$ | Carrier/dopant ** | 38 |

# References

1. Wang, L.; Gutiérrez-Lezama, I.; Barreteau, C.; Ubrig, N.; Giannini, E.; Morpurgo, A. F., Tuning magnetotransport in a compensated semimetal at the atomic scale. *Nature Communications* **2015,** *6*, 8892.
2. Liu, W. L.; Chen, M. L.; Li, X. X.; Dubey, S.; Xiong, T.; Dai, Z. M.; Yin, J.; Guo, W. L.; Ma, J. L.; Chen, Y. N.; Tan, J.; Li, D.; Wang, Z. H.; Li, W.; Bouchiat, V.; Sun, D. M.; Han, Z.; Zhang, Z. D., Effect of aging-induced disorder on the quantum transport properties of few-layer WTe2. *2D Materials* **2016,** *4* (1), 011011.
3. Ali, M. N.; Xiong, J.; Flynn, S.; Tao, J.; Gibson, Q. D.; Schoop, L. M.; Liang, T.; Haldolaarachchige, N.; Hirschberger, M.; Ong, N. P.; Cava, R. J., Large, non-saturating magnetoresistance in WTe2. *Nature* **2014,** *514* (7521), 205-+.
4. Zhao, Y.; Liu, H.; Yan, J.; An, W.; Liu, J.; Zhang, X.; Wang, H.; Liu, Y.; Jiang, H.; Li, Q.; Wang, Y.; Li, X.-Z.; Mandrus, D.; Xie, X. C.; Pan, M.; Wang, J., Anisotropic magnetotransport and exotic longitudinal linear magnetoresistance in $\mathrm{WT}{\mathrm{e}}_{2}$ crystals. *Physical Review B* **2015,** *92* (4), 041104.
5. Pan, X.-C.; Pan, Y.; Jiang, J.; Zuo, H.; Liu, H.; Chen, X.; Wei, Z.; Zhang, S.; Wang, Z.; Wan, X.; Yang, Z.; Feng, D.; Xia, Z.; Li, L.; Song, F.; Wang, B.; Zhang, Y.; Wang, G., Carrier balance and linear magnetoresistance in type-II Weyl semimetal WTe2. *Front Phys-Beijing* **2017,** *12* (3), 127203.
6. Wang, Y.; Liu, E.; Liu, H.; Pan, Y.; Zhang, L.; Zeng, J.; Fu, Y.; Wang, M.; Xu, K.; Huang, Z.; Wang, Z.; Lu, H.-Z.; Xing, D.; Wang, B.; Wan, X.; Miao, F., Gate-tunable negative longitudinal magnetoresistance in the predicted type-II Weyl semimetal WTe2. *Nature Communications* **2016,** *7* (1), 13142.
7. Li, P.; Wen, Y.; He, X.; Zhang, Q.; Xia, C.; Yu, Z. M.; Yang, S. Y. A.; Zhu, Z. Y.; Alshareef, H. N.; Zhang, X. X., Evidence for topological type-II Weyl semimetal WTe2. *Nature Communications* **2017,** *8*.
8. Dai, X.; Lu, H.-Z.; Shen, S.-Q.; Yao, H., Detecting monopole charge in Weyl semimetals via quantum interference transport. *Physical Review B* **2016,** *93* (16), 161110.
9. Lee, P. A.; Ramakrishnan, T. V., Disordered electronic systems. *Reviews of Modern Physics* **1985,** *57* (2), 287-337.
10. Lu, H.-Z.; Shen, S.-Q., Weak antilocalization and localization in disordered and interacting Weyl semimetals. *Physical Review B* **2015,** *92* (3), 035203.
11. Sangiao, S.; Morellon, L.; Simon, G.; De Teresa, J. M.; Pardo, J. A.; Arbiol, J.; Ibarra, M. R., Anomalous Hall effect in Fe (001) epitaxial thin films over a wide range in conductivity. *Physical Review B* **2009,** *79* (1), 014431.
12. Mott, N. F., Conduction in non-crystalline materials. *The Philosophical Magazine: A Journal of Theoretical Experimental and Applied Physics* **1969,** *19* (160), 835-852.
13. Mleczko, M. J.; Xu, R. L.; Okabe, K.; Kuo, H.-H.; Fisher, I. R.; Wong, H. S. P.; Nishi, Y.; Pop, E., High Current Density and Low Thermal Conductivity of Atomically Thin Semimetallic WTe2. *ACS Nano* **2016,** *10* (8), 7507-7514.
14. Liu, L.; Moriyama, T.; Ralph, D. C.; Buhrman, R. A., Spin-Torque Ferromagnetic Resonance Induced by the Spin Hall Effect. *Physical Review Letters* **2011,** *106* (3).

15. Chang, H.; Li, P.; Zhang, W.; Liu, T.; Hoffmann, A.; Deng, L.; Wu, M., Nanometer-Thick Yttrium Iron Garnet Films With Extremely Low Damping. *IEEE Magnetics Letters* **2014,** *5*, 1-4.
16. Kumar, A.; Akansel, S.; Stopfel, H.; Fazlali, M.; Åkerman, J.; Brucas, R.; Svedlindh, P., Spin transfer torque ferromagnetic resonance induced spin pumping in the Fe/Pd bilayer system. *Physical Review B* **2017,** *95* (6), 064406.
17. Zhu, L.; Ralph, D. C.; Buhrman, R. A., Effective Spin-Mixing Conductance of Heavy-Metal--Ferromagnet Interfaces. *Physical Review Letters* **2019,** *123* (5), 057203.
18. Zhao, B.; Khokhriakov, D.; Zhang, Y.; Fu, H.; Karpiak, B.; Hoque, A. M.; Xu, X.; Jiang, Y.; Yan, B.; Dash, S. P., Observation of charge to spin conversion in Weyl semimetal ${\mathrm{WTe}}_{2}$ at room temperature. *Physical Review Research* **2020,** *2* (1), 013286.
19. Liu, J.; Ohkubo, T.; Mitani, S.; Hono, K.; Hayashi, M., Correlation between the spin Hall angle and the structural phases of early 5d transition metals. *Applied Physics Letters* **2015,** *107* (23), 232408.
20. Zhu, L.; Zhu, L.; Sui, M.; Ralph, D. C.; Buhrman, R. A., Variation of the giant intrinsic spin Hall conductivity of Pt with carrier lifetime. *Science Advances* **2019,** *5* (7), eaav8025.
21. Shi, S.; Liang, S.; Zhu, Z.; Cai, K.; Pollard, S. D.; Wang, Y.; Wang, J.; Wang, Q.; He, P.; Yu, J.; Eda, G.; Liang, G.; Yang, H., All-electric magnetization switching and Dzyaloshinskii-Moriya interaction in WTe2/ferromagnet heterostructures. *Nat Nanotechnol* **2019,** *14* (10), 945-949.
22. Fan, Y.; Upadhyaya, P.; Kou, X.; Lang, M.; Takei, S.; Wang, Z.; Tang, J.; He, L.; Chang, L. T.; Montazeri, M.; Yu, G.; Jiang, W.; Nie, T.; Schwartz, R. N.; Tserkovnyak, Y.; Wang, K. L., Magnetization switching through giant spin-orbit torque in a magnetically doped topological insulator heterostructure. *Nat Mater* **2014,** *13* (7), 699-704.
23. Hayashi, M.; Kim, J.; Yamanouchi, M.; Ohno, H., Quantitative characterization of the spin-orbit torque using harmonic Hall voltage measurements. *Physical Review B* **2014,** *89* (14), 144425.
24. Avci, C. O.; Garello, K.; Gabureac, M.; Ghosh, A.; Fuhrer, A.; Alvarado, S. F.; Gambardella, P., Interplay of spin-orbit torque and thermoelectric effects in ferromagnet/normal-metal bilayers. *Physical Review B* **2014,** *90* (22), 224427.
25. Liu, L.; Pai, C. F.; Li, Y.; Tseng, H. W.; Ralph, D. C.; Buhrman, R. A., Spin-torque switching with the giant spin Hall effect of tantalum. *Science* **2012,** *336* (6081), 555-8.
26. Chen, T.-Y.; Chan, H.-I.; Liao, W.-B.; Pai, C.-F., Current-Induced Spin-Orbit Torque and Field-Free Switching in $\mathrm{Mo}$-Based Magnetic Heterostructures. *Physical Review Applied* **2018,** *10* (4), 044038.
27. Li, P.; Wu, W. K.; Wen, Y.; Zhang, C. H.; Zhang, J. W.; Zhang, S. F.; Yu, Z. M.; Yang, S. Y. A.; Manchon, A.; Zhang, X. X., Spin-momentum locking and spin-orbit torques in magnetic nano-heterojunctions composed of Weyl semimetal WTe2. *Nature Communications* **2018,** *9*.
28. Aradhya, S. V.; Rowlands, G. E.; Oh, J.; Ralph, D. C.; Buhrman, R. A., Nanosecond-Timescale Low Energy Switching of In-Plane Magnetic Tunnel Junctions through Dynamic Oersted-Field-Assisted Spin Hall Effect. *Nano Letters* **2016,** *16* (10), 5987-5992.
29. MacNeill, D.; Stiehl, G. M.; Guimarães, M. H. D.; Reynolds, N. D.; Buhrman, R. A.; Ralph, D. C., Thickness dependence of spin-orbit torques generated by ${\text{WTe}}_{2}$. *Physical Review B* **2017,** *96* (5), 054450.

30. Lv, Y.; Kally, J.; Zhang, D.; Lee, J. S.; Jamali, M.; Samarth, N.; Wang, J.-P., Unidirectional spin-Hall and Rashba–Edelstein magnetoresistance in topological insulator-ferromagnet layer heterostructures. *Nature Communications* **2018,** *9* (1), 111.
31. Avci, C. O.; Garello, K.; Ghosh, A.; Gabureac, M.; Alvarado, S. F.; Gambardella, P., Unidirectional spin Hall magnetoresistance in ferromagnet/normal metal bilayers. *Nature Physics* **2015,** *11*, 570.
32. Avci, C. O.; Mendil, J.; Beach, G. S. D.; Gambardella, P., Origins of the Unidirectional Spin Hall Magnetoresistance in Metallic Bilayers. *Physical Review Letters* **2018,** *121* (8), 087207.
33. Avci, C. O.; Garello, K.; Mendil, J.; Ghosh, A.; Blasakis, N.; Gabureac, M.; Trassin, M.; Fiebig, M.; Gambardella, P., Magnetoresistance of heavy and light metal/ferromagnet bilayers. *Applied Physics Letters* **2015,** *107* (19), 192405.
34. Avci, C. O.; Mann, M.; Tan, A. J.; Gambardella, P.; Beach, G. S. D., A multi-state memory device based on the unidirectional spin Hall magnetoresistance. *Applied Physics Letters* **2017,** *110* (20).
35. Lv, Y.; Kally, J.; Liu, T.; Sahu, P.; Wu, M.; Samarth, N.; Wang, J.-P., Large unidirectional spin Hall and Rashba-Edelstein magnetoresistance in topological insulator/magnetic insulator heterostructures. *arXiv preprint arXiv:1806.09066* **2018**.
36. Olejník, K.; Novák, V.; Wunderlich, J.; Jungwirth, T., Electrical detection of magnetization reversal without auxiliary magnets. *Physical Review B* **2015,** *91* (18), 180402.
37. Yasuda, K.; Tsukazaki, A.; Yoshimi, R.; Takahashi, K. S.; Kawasaki, M.; Tokura, Y., Large Unidirectional Magnetoresistance in a Magnetic Topological Insulator. *Physical Review Letters* **2016,** *117* (12), 127202.
38. Fan, Y. B.; Shao, Q. M.; Pan, L.; Che, X. Y.; He, Q. L.; Yin, G.; Zheng, C.; Yu, G. Q.; Nie, T. X.; Masir, M. R.; MacDonald, A. H.; Wang, K. L., Unidirectional Magneto-Resistance in Modulation-Doped Magnetic Topological Insulators. *Nano Letters* **2019,** *19* (2), 692-698.
39. MacNeill, D.; Stiehl, G. M.; Guimaraes, M. H. D.; Buhrman, R. A.; Park, J.; Ralph, D. C., Control of spin–orbit torques through crystal symmetry in WTe2/ferromagnet bilayers. *Nature Physics* **2016,** *13*, 300.